\documentclass[a4paper]{article}
\usepackage{amsfonts}
\usepackage{amsmath}
\usepackage{physics}
\usepackage{jheppub}
\usepackage{lmodern}
\usepackage{graphicx}
\usepackage[section]{placeins}
\usepackage{xcolor}
\usepackage[utf8]{inputenc}
\usepackage[T1]{fontenc}
\usepackage{lmodern}
\definecolor{refkey}{gray}{0.45}
\definecolor{labelkey}{RGB}{155,48,48}
\hypersetup{
  colorlinks=true,
  citecolor=magenta,
  linkcolor=blue,
  urlcolor=violet
 }

\def\be{\begin{equation}}\def\ee{\end{equation}}
\def\del{\partial}
\def\md{\mathrm{d}}
\def\mi{\mathrm{i}}

\title{Jackiw-Teitelboim Gravity and Rotating Black Holes}

\author{Upamanyu Moitra,}
\author{Sunil Kumar Sake,}
\author{Sandip P. Trivedi,}
\author{V. Vishal}

\affiliation{Department of Theoretical Physics, Tata Institute of Fundamental Research, Mumbai -- 400005, India}

\emailAdd{upamanyu@theory.tifr.res.in}
\emailAdd{sunil.sake@tifr.res.in}
\emailAdd{sandip@theory.tifr.res.in}
\emailAdd{vishal@theory.tifr.res.in}

\preprint{\parbox{3cm}{TIFR/TH/19-17}}

\abstract{
We show that the free energy at low temperatures for  near-extremal black holes is correctly obtained from the Jackiw-Teitelboim (JT) model of gravity. 
Our arguments apply to all black holes, including rotating ones, whose metric has a near-horizon $\mathrm{AdS}_2$ factor and the associated $\mathrm{SL}(2,\mathbb{R})$ symmetry. 
We  verify these arguments by explicit calculations for  rotating black holes in $4$ and $5$ dimensions.   Our results suggest  that the JT model could prove useful in analysing the dynamics of near-extremal Kerr black holes found in nature. 
}

\begin{document}

\maketitle
\flushbottom

\section{Introduction}

\label{sec-intro}
Exciting recent developments have revealed that the behaviour of a class of models studied in condensed matter physics, which we will refer to generically as Sachdev-Ye-Kitaev (SYK) models here
\cite{Sachdev:1992fk,Kitaev}, is connected with the behaviour of near-extremal black holes. 

More precisely, it was shown in \cite{Maldacena:2016upp} that the behaviour of the Jackiw-Teitelboim (JT) theory \cite{Teitelboim:1983ux,Jackiw:1984je,Almheiri:2014cka} of $2$-dimensional gravity has many parallels with the SYK model.
In particular, in the SYK model, conformal invariance is broken, and this breaking determines the low temperature thermodynamics and low-energy response of the system. 
Similarly, conformal invariance is also broken  in the JT model, and this breaking again determines the   thermodynamics and response of the system.

In subsequent works,  \cite{Nayak:2018qej, Moitra:2018jqs}  it was shown that in fact the JT model is a good approximation for the dynamics and thermodynamics  of a large class of charged black holes close to extremality.
 And by understanding the breaking of conformal invariance, a precise version of the near-$\mathrm{AdS}_2$/near-$\mathrm{CFT}_1$ correspondence was also developed which applies to these black holes. The only requirement for these considerations to apply  is that the near-horizon geometry at extremality has an $\mathrm{AdS}_2$ factor and the $\mathrm{SL}(2,\mathbb{R})$ symmetry associated with it. 
 This is true for  the Reissner-Nordstr\"{o}m black hole, and several other black holes which can arise in string theory, including those  in systems which have extra scalars and gauge fields.
 
 In this paper, we extend these considerations to rotating black holes. Several examples of extremal rotating  black holes including the extremal Kerr black hole in flat or asymptotically $\mathrm{AdS}$ spacetime are known to posses an $\mathrm{AdS}_2$ factor and an $\mathrm{SL}(2,\mathbb{R})$ symmetry in their near-horizon geometry. 
 We show that the low-temperature thermodynamics, close to extremality, for these black holes is correctly obtained from a JT model with appropriate values for the two-dimensional Newton's constant and the scale of conformal symmetry breaking. This is the central result of the paper. 
 
 The paper is structured as follows. 
 After a review of the connection between the JT model and near-extremal Reissner-Nordstr\"{o}m black holes in section \ref{sec-rev-sph}, we give an explanation for why the JT model should give the correct low temperature free energy and associated thermodynamics for all spherically symmetric near-extremal black holes with a near-horizon $\mathrm{AdS}_2$ geometry
 in \S\ref{ssec-mdrn}. These arguments are extended to the rotating case in general in \S\ref{ss-jtrot}, once again for all cases where the extremal geometry has an $\mathrm{AdS}_2$ near-horizon geometry.
 
In the following two sections, \ref{sec5drf} and \ref{sec-4drbf}, we consider several examples, including the near-extremal Kerr black hole in $5$ dimensions  in asymptotically $\mathrm{AdS}$ spacetime, the near-extremal dyonic Kerr-Newman black hole in asymptotically $\mathrm{AdS}$ spacetime in $4$ dimensions, and in particular, the near-extremal Kerr black hole in $4$ dimensions in asymptotically flat spacetime.  By explicit calculations, we
  show that the general conclusion referred to above is correct, namely the low temperature  behaviour of their free energy agrees with that of the JT model. 
 
 Our results about the wide applicability of the JT model for the rotating case as well, suggests, in fact, that this model could be useful for studying some aspects of the dynamics of near-extremal Kerr black holes  found in nature. We end the paper by making some comments along these lines, along with a summary and some conclusions in section \ref{sec-cocl}. 
 
 Appendices \ref{app-jttd}--\ref{app-rot4d} contain some important additional supplementary material.

 Before we proceed, let us mention some of the relevant literature. The various aspects of the SYK model, its variants and  the JT model have also  been studied in, e.g., 
 \cite{Muta:1992xw, Lemos:1993qn, Lemos:1996bq, Polchinski:2016xgd, Jevicki:2016bwu, Maldacena:2016hyu, Jensen:2016pah, Danshita:2016xbo, Engelsoy:2016xyb, Almheiri:2016fws, Bagrets:2016cdf, Cvetic:2016eiv, Jevicki:2016ito, Gu:2016oyy, Gross:2016kjj, Berkooz:2016cvq,  Garcia-Garcia:2016mno, Fu:2016vas, Witten:2016iux, Gurau:2016lzk, Cotler:2016fpe, Klebanov:2016xxf, Davison:2016ngz, Peng:2016mxj, Krishnan:2016bvg, Turiaci:2017zwd, Li:2017hdt, Gurau:2017xhf, Mandal:2017thl, Bonzom:2017pqs, Gross:2017hcz, Stanford:2017thb, Krishnan:2017ztz, Maldacena:2017axo, Das:2017pif, Narayan:2017qtw, Chaudhuri:2017vrv, Mertens:2017mtv, Murugan:2017eto, Krishnan:2017txw, Gross:2017vhb, Eberlein:2017wah, Taylor:2017dly, Garcia-Garcia:2017bkg, Kourkoulou:2017zaj,  Anninos:2017cnw, Giombi:2017dtl, Sonner:2017hxc, Bulycheva:2017ilt, Choudhury:2017tax, Grumiller:2017qao, Gross:2017aos, Kitaev:2017awl, Das:2017hrt, Narayan:2017hvh, Das:2017wae, Gonzalez:2018enk, Krishnan:2018hhu, Roberts:2018mnp, Benedetti:2018goh, Gaikwad:2018dfc, Klebanov:2018nfp, Kolekar:2018sba, Gharibyan:2018jrp, Maldacena:2018lmt, Harlow:2018tqv, Brown:2018kvn, Lam:2018pvp, Bena:2018bbd, Saad:2018bqo, Gubser:2018yec, Larsen:2018iou, Blommaert:2018oro, Chang:2018sve, Gur-Ari:2018okm, Goel:2018ubv, Lin:2018xkj, Castro:2018ffi, Liu:2018jhs, Giombi:2018qgp, Kitaev:2018wpr, Pakrouski:2018jcc, Blake:2018leo, Yang:2018gdb, Brown:2018bms, Kolekar:2018chf, Larsen:2018cts, Bhattacharya:2018fkq, Alishahiha:2018swh, Dhar:2018pii, Murugan:2018fdj, Goto:2018iay, Kim:2019upg, Sachdev:2019bjn, Blommaert:2019hjr, Nayak:2019khe, Sun:2019mms, Jahnke:2019gxr, Saad:2019lba, Mertens:2019bvy, Maldacena:2019cbz, Guo:2019csw, Mertens:2019tcm, Susskind:2019ddc, Lin:2019qwu, Iliesiu:2019xuh, Cotler:2019nbi, Klebanov:2019jup, Sun:2019yqp}. For a pedagogical review of the SYK model, see \cite{Rosenhaus:2018dtp}; the connections with $\mathrm{AdS}_2$ holography are reviewed in  \cite{Sarosi:2017ykf}. The Kerr-CFT correspondence was studied earlier in  \cite{Guica:2008mu} and was further explored in, for instance,  \cite{Hartman:2008pb, Hartman:2009nz, Bredberg:2009pv, Guica:2010ej,  Castro:2010fd, Porfyriadis:2014fja, Hadar:2014dpa, Hadar:2015xpa, Porfyriadis:2016gwb}; for a review, see \cite{Compere:2012jk}.

\section{A Review of the Jackiw-Teitelboim Model and Near-Extremal Reissner-Nordstr\"{o}m Black Holes}
\label{sec-rev-sph}
\subsection{Jackiw-Teitelboim Model}

Let us briefly recapitulate some of the key features of the Jackiw-Teitelboim (JT) model  of two-dimensional gravity and how it describes near-extremal Reissner-Nordstr\"{o}m (RN) black holes. 
The JT model consists of two-dimensional gravity with a dilaton $\phi$ described by the action:
\be
\label{JTact}
I= -{1\over {16 \pi \widetilde G}}  \left(  \int \md^2 x \, \sqrt{g} \, \phi(R-\Lambda_{2} ) + 2 \int_{\del} \md x \, \sqrt{\gamma} \, \phi K  \right).
\ee

Note that we are working in Euclidean spacetime here since this paper is mostly concerned with black hole thermodynamics. 

The two-dimensional spacetimes one considers in the model have a one-dimensional boundary, denoted by $\del$ in eq.(\ref{JTact}),  at a fixed  value of the dilaton
\be
\label{bcdil}
\phi=\phi_B.
\ee
In addition, fluctuations in the  metric also vanish sufficiently fast  at the boundary; we will be more precise about this shortly. 

The equations of motion then lead to the  geometry being  $\mathrm{AdS}_2$ 
\be
\label{metads2}
\md s^2={L_2^2 \over z^2}(\md t^2+\md z^2),
\ee
with radius $L_2$ being  given by, 
\be
\label{JTdefL2} 
L_2 = \sqrt{- \frac{2}{\Lambda_{2 } }}.
\ee
In addition, the dilaton can be linearly varying 
\be
\label{linvardil}
\phi={1\over \mathcal{J} z}.
\ee
This linear variation breaks the scale invariance of $\mathrm{AdS}_2$ and is parametrised by the scale $\mathcal{J}$ which has dimensions of $ [M]\sim [L]^{-1}$.
This breaking of scale invariance is very important in the resulting behaviour of the system. 

The boundary conditions imposed on the metric can be now stated more precisely. In a coordinate system with $g_{tz}=0$, 
\begin{align}
g_{tt} & \rightarrow  {L_2^2\over z^2} \left( 1+ \mathcal{O}(z^2) \right),  \nonumber \\
g_{zz} & \rightarrow  {L_2^2\over z^2}\left( 1+ \mathcal{O}(z^2) \right), \label{condmet}
\end{align}
as $z\rightarrow z_B$, where $\phi=\phi_B$.
As was discussed in \cite{Nayak:2018qej}, these boundary conditions, along with the Dirichlet boundary condition for the dilaton, eq.(\ref{bcdil}), give rise to a well-defined variational principle,
with the equations of motion having a solution, eqs.(\ref{metads2}), (\ref{linvardil}).

In our discussion below, where the JT model arises from black holes in higher dimensions, $z_B$ will be small  enough at the boundary so that the $\mathcal{O}(z^2)$ and higher corrections in eq.(\ref{condmet}) are suppressed, but not too small, so that, eq.(\ref{linvardil}), 
\be
\label{condphib}
\phi_B ={1\over \mathcal{J} z_B}
\ee
  is also small meeting the condition, $\phi_B \ll 1$. 

The dynamics of the system arise from fluctuations of the boundary which can be parametrised by time reparametrisations and is  described by a Schwarzian action,
\be 
\label{act-schw}
I = -{1 \over 8\pi \widetilde{G} \mathcal{J} } \int \md \tau \, \mathrm{Sch} (t, \tau),
\ee
where,
\be 
\label{def-sch}
\mathrm{Sch} (t, \tau) \equiv \frac{t'''(\tau)}{t'(\tau)} - \frac{3}{2} \left( \frac{t''(\tau)}{t'(\tau)} \right)^2.
\ee
Here $t$ is the Poincar\'e time coordinate in eq.(\ref{metads2}) and $\tau$ is the proper time along the boundary defined by,
\be 
\label{jt-def-tau}
\frac{\md \tau^2}{z_B^2} = \frac{1}{z^2} (\md t^2 + \md z^2) \Big|_{\del},
\ee
where $z_B$ is location of the boundary, eq.(\ref{bcdil}) in the $z$ coordinate.

The reader will notice that in the action, eq.(\ref{JTact}), we have included a Gibbons-Hawking (GH)  boundary term,
\be
\label{GHJT}
I_{\mathrm{GH}}=-{1\over 8 \pi {\widetilde G} } \int_{\del} \md x \, \sqrt{\gamma} \, \phi K. 
\ee

In addition to the action eq.(\ref{JTact}),  we also need to add a counter-term,
\be
\label{jt-ct}
I_{\mathrm{CT} } =  \frac{1}{8 \pi \widetilde{G} L_2} \int_{\del } \md x\, \sqrt{\gamma}\, \phi, 
\ee
which cancels the divergence which arises from the GH term in evaluating the on-shell action, to obtain a finite mass and a finite on-shell action. 
The ADM mass is then given by
\be 
\label{JT-adm}
M_{\mathrm{ADM}} = \frac{1}{8\pi \widetilde{G} \mathcal{J}} \mathrm{Sch} (t, \tau),
\ee
and is also  proportional to the Schwarzian. 
The free energy  at temperature $T$ of the system is given by
\be
\label{fJT}
F= - \frac{\pi}{4\widetilde{G} \mathcal{J} } T^2,
\ee 
and is quadratic in $T$. Notice that the scale $\mathcal{J}$ appears in the denominator in the Schwarzian action, mass and free energy. 

\subsection{Near-Extremal  Reissner-Nordstr\"{o}m Black Holes}
Next consider a near-extremal RN geometry in $4$ dimensions in either asymptotically flat or $\mathrm{AdS}_4$ spacetime with metric
\be
\label{neRN}
\md s^2=f(r)\, \md t^2+{\md r^2\over f(r)} + \Phi^2(r) \md\Omega_{2}^2.
\ee
Here, 
\be
\label{valf}
f(r)= \left(1 - \frac{2G_{4} M}{r} + \frac{Q^2}{r^2} + \frac{r^2}{L^2} \right),
\ee
and
\be
\label{dilern}
\Phi(r)=r,
\ee
$L$ is the radius of $\mathrm{AdS}_4$ and the flat spacetime limit is obtained by taking $L\rightarrow \infty$. 
For an electrically charged black hole, the electromagnetic field strength is 
\be
\label{elec}
F_{rt}={Q\over r^2}.
\ee

In the extremal limit,  $f(r)$ has a ``double zero''  at  a radial location  we denote by $r_h$ and is given by
\be
\label{bef}
f_0 (r) = { (r-r_h)^2 \over L_2^2}
\ee 
in the vicinity of the horizon. The radial location of the horizon, $r_h$,  is an important parameter in the subsequent discussion.  It determines the radius of the horizon, eq.(\ref{dilern}) and is  fixed  by $Q$ and in the  asymptotically $\mathrm{AdS}$ case, also by 
the radius of $\mathrm{AdS}_4$, $L$. The near-horizon geometry for an extremal RN black hole is $\mathrm{AdS}_2\times S^2$ with the radius $L_2$, eq.(\ref{metads2}), also  being determined in general by the charge $Q$,
and $L$. The coordinate $z$ in eq.(\ref{metads2}) is related to the radial coordinate in eq.(\ref{neRN}) by 
\be
\label{coordt}
z={L_2^2\over (r-r_h)}.
\ee

The spherically symmetric geometry in  eq.(\ref{neRN})  is suggestive and we can carry out the  dimensional reduction over the $S^2$ and construct 
a $2$-dimensional system consisting of gravity and a scalar $\Phi$, which is the radius of the two-sphere.  This two-dimensional theory   turns out to be well
approximated by the JT model which  accurately describes the thermodynamics at low temperatures \cite{Nayak:2018qej}.  We review how this connection with the JT model arises next. Before proceeding, let us  note that for the electrically charged case we discuss below an additional phase mode is also present. Its action can also be expressed in terms of a term on the boundary, 
\be
\label{actphase}
I_{\mathrm{phase}}=  - \mathcal{I}   \int_{\del} \md \tau \, ({\dot \theta})^2.
\ee
How this mode arises and how the resulting value of the coefficient, $\mathcal{I}$, which is related to the charge susceptibility  \cite{Davison:2016ngz}, is   obtained was discussed in some detail in  \cite{Moitra:2018jqs}.  We mention it only in passing here since 
we will mainly study the free energy at fixed charge in this paper and the phase mode is not very relevant in the resulting   analysis.

We start with the $4$-dimensional theory described by the Euclidean action,
\begin{equation}
\label{maxact}
I = -\frac{1}{16 \pi G_{4} } \left( \int \md^4 x\,\sqrt{g} \,( R - 2 \Lambda_4 + F_{\mu \nu} F^{\mu \nu} ) + 2 \int \md^3 x \, \sqrt{\gamma} \, K    \right).
\end{equation}

Taking the metric of the form
\be
\label{met2}
\md s^2= {g}_{\alpha \beta} \, \md x^\alpha \md x^\beta + \Phi^2 \md\Omega_{2}^2,
\ee
and dimensionally reducing over the $S^2$ (here the field $\Phi$ and the metric depend only on the 2 coordinates $x^\alpha$) and further changing frames under the transformation
\be 
\label{jt-weyl}
{g}_{\alpha \beta} \rightarrow \frac{\Phi_0}{\Phi} g_{\alpha \beta},
\ee
gives the two-dimensional action,
\be
\label{dimredb}
I= - \frac{1}{4G_{4} } \int \md^2 x \, \sqrt{g} \, \left(  \Phi^2 R + \frac{2\Phi_0}{\Phi} - 2\Phi_0 \Phi \Lambda_4 + \frac{\Phi^3}{\Phi_0} F^{\alpha \beta} F_{\alpha \beta}  \right) - \frac{1}{2G_{4} } \int \md x \, \sqrt{\gamma} \, \Phi^2 K.
\ee

This action has a solution where the metric is $\mathrm{AdS}_2$, $\Phi$ takes the attractor value 
\be
\label{defphio}
\Phi_0=r_h=  { L \over \sqrt{6} } \left( \sqrt{1+ \frac{12 Q^2}{L^2}} - 1 \right)^{1/2},
\ee
 and the field strength is given by 
\be
\label{fs}
F_{\alpha \beta} \, \md x^\alpha \wedge \md x^\beta ={Q \Phi_0 \over \Phi^3} \sqrt{g} \, \epsilon_{\alpha \beta} \,  \md x^\alpha \wedge \md x^\beta,
\ee
where $\epsilon_{\alpha \beta}$ is the Levi-Civita symbol (anti-symmetric tensor density). Note that throughout this paper, we have chosen the convention that in the $r,t$ coordinates, $\epsilon_{rt}  = 1$.

The on-shell value of the action $I$ is known to give  the grand canonical partition function. As was mentioned above, here we will consider the canonical partition function. This is obtained by adding a boundary term to $I$ \cite{Hawking:1995ap},
\be
\label{bdgf}
I_{\del \mathrm{GF}} = \frac{1}{G_{4} } \int \md x\, \sqrt{\gamma} \, \frac{\Phi^3}{\Phi_0} n_\alpha F^{\alpha \beta} A_\beta.
\ee
in the resulting two-dimensional theory, eq.(\ref{dimredb}).  Here $n_\alpha$ is the outward normal one-form on the boundary. After including it, we get, 
\begin{align}
I_{\mathrm{canonical}} &= - \frac{1}{4G_{4} } \int \md^2 x \, \sqrt{g} \left(  \Phi^2 R + \frac{2\Phi_0}{\Phi} - 2\Phi_0 \Phi \Lambda_4 + \frac{\Phi^3}{\Phi_0} F^{\alpha \beta} F_{\alpha \beta}  \right) \nonumber \\  
&\quad - \frac{1}{2G_{4} } \int \md x \, \sqrt{\gamma}\, \left( \Phi^2 K - \frac{2\Phi^3}{\Phi_0} n_\alpha F^{\alpha \beta} A_\beta \right). \label{dimreda}
\end{align}

Expanding about the attractor value for $\Phi$, eq.(\ref{defphio}),   
\be
\label{defphi}
\Phi=\Phi_0(1+\phi),
\ee
 and inserting in eq.(\ref{dimreda}) then gives
\begin{align}
\label{act3}
I & =   - \frac{r_h^2}{4 G_{4} } \left( \int \md^2 x\, \sqrt{g} \,  R + 2 \int \md x\, \sqrt{\gamma} \, K  \right) \cr
&\quad - \frac{r_h^2}{2G_{4} } \left(  \int \md^2 x\, \sqrt{g} \, \phi \Big(R + \frac{2}{L_2^2}\Big) + 2 \int \md x\, \sqrt{\gamma} \, \phi  K \right)  \cr
&\quad  - \frac{r_h^2}{4G_{4} } \left(  \int \md^2 x\, \sqrt{g} \, \phi^2 \Big(R - \frac{6}{L_2^2} - \frac{4}{r_h^2}\Big) + 2 \int \md x\, \sqrt{\gamma} \, \phi^2  K \right) + \cdots .
\end{align}

On the RHS above, the first line, which is independent of $\phi$, gives the entropy  for the extremal black hole. 
For computing the free energy above extremality, 
\be
\label{defdF}
\Delta F = F-F_{\mathrm{extremal}}
\ee
 as a function of temperature, we can neglect terms in this line.  The last line in eq.(\ref{act3}) has corrections which are quadratic in $\phi$, as well as additional terms indicated by the ellipsis  which include  terms  with even higher orders in $\phi$. We will also see in the next section why these terms  can be neglected for the leading order temperature dependence. This leaves the second line in eq.(\ref{act3}) which is exactly the action for the  JT model.
Comparing with eq.(\ref{JTact}) we also see that the resulting value of the two-dimensional Newton's constant is, 
\be
\label{valgt}
 {\widetilde G}= {G_{4} \over 8\pi r_h^2}.
\ee
In addition to the terms in eq.(\ref{act3}), we note that the JT model also has the counter-term, eq.(\ref{jt-ct}). We also note that the additional boundary term we added,
eq.(\ref{bdgf}), when expanded at linear order in $\phi$ becomes, 
\be
\label{bgfl}
I_{\del \mathrm{GF}} =  \frac{r_h^2}{G_{4} }  \int \md x\, \sqrt{\gamma} \, (1+3\phi) n_\alpha F^{\alpha \beta} A_\beta.
\ee

We note that if we treated (\ref{act3}) as a two-dimensional action in its own right, including the terms of higher order in $\phi$, the remarkable simplification accorded by the JT model would be lost. However, one can still make progress in principle, using, for example, a perturbative approach -- in which the corrections can be kept systematically.

To make the connection with the JT model more precise, we need to be  clearer about where the boundary of spacetime is located. To begin, the spacetime has a boundary at asymptotic infinity; it is at this boundary that the GH term  and the additional boundary term involving the gauge field in eq.(\ref{dimreda}) are to be evaluated. However, in the JT model, the boundary is not at asymptotic infinity. It is located instead in the ``asymptotic $\mathrm{AdS}_2$'' region. The two boundary terms, the GH term and the one involving the gauge field are evaluated at this boundary to obtain the action, eq.(\ref{act3}). 

Let us now explain where the asymptotic $\mathrm{AdS}_2$  region lies in spacetime. 
From eq.(\ref{dilern}) and eq.(\ref{defphi}), we see that 
\be
\label{valphia}
\phi={r-r_h\over r_h}={L_2^2\over r_h} {1\over z}.
\ee
Comparing with eq.(\ref{linvardil}) we see therefore that the energy scale which characterises the breaking of scale invariance in the near-horizon region is 
\be
\label{valJ}
\mathcal{J}={r_h\over L_2^2}.
\ee
We are   interested  here in the behaviour at sufficiently small temperatures. At these low  temperatures, the geometry has a region, which is sufficiently far from the horizon so that  the effects of finite temperature have ``died down'', but not so far that the effects of the breaking of scale invariance due to the scale $\mathcal{J}$ have become significant. This is the asymptotic $\mathrm{AdS}_2$ region we referred to above. 
The boundary in  the JT model we consider will be located in this asymptotic $\mathrm{AdS}_2$ region at a fixed value of $\phi$, eq.(\ref{bcdil}).

We can now be more precise about how low $T$ should be for our considerations to apply. We take  $T$ to satisfy the condition,  
\be
\label{valT}
T\ll \mathcal{J}.
\ee
The $\mathcal{O}(z^2)$ corrections in eq.(\ref{condmet}) for the black hole solutions are actually of order $(Tz)^2$. The asymptotic $\mathrm{AdS}_2$ region then lies in the range where 
\be
\label{valzb}
{1\over \mathcal{J} }\ll z\ll {1\over T}.
\ee
The boundary  which lies at $z_B$ also satisfies this condition.
It follows  then from  the upper end of the  inequality in eq.(\ref{valzb}) 
that in this region  the $\mathcal{O}(z^2)$ corrections to the metric are small, while it follows from the lower end   that  the boundary  is still in the near-horizon region far from the asymptotically flat or $\mathrm{AdS}_4$ region of spacetime. These considerations were also discussed above around eq.(\ref{condphib}).  It is worth emphasising that  our considerations apply to near-extremal black holes in asymptotically $\mathrm{AdS}$ and flat spacetimes, with the boundary of the $\mathrm{AdS}_2$ region satisfying eq.(\ref{valzb}). The relation between the scale $\mathcal{J}$ and the underlying parameters varies depending on the kind of black hole being considered. For related comments, see \S\ref{ss-jtsc} of this paper and ref.\cite{Moitra:2018jqs}.

Note that from eq.(\ref{valphia}) and eq.(\ref{valzb}) it also follows that 
\be
\label{condpc}
\phi \ll 1
\ee
in the region of spacetime extending from the horizon  to the asymptotic $\mathrm{AdS_2}$ boundary, see eq.(\ref{condphib}) and the comment thereafter. Away from the $\mathrm{AdS_2}$ boundary, $\phi$ no longer satisfies the condition eq.(\ref{condpc}) and  becomes  large towards asymptotic infinity.

Comparing  with the free energy for  the JT model,  eq.(\ref{fJT}), and keeping track of the values for ${\widetilde G}$, and $\mathcal{J}$, eq.(\ref{valgt}) and (\ref{valJ}), we then get that the temperature dependence of the free energy  for a near-extremal RN black hole is given by 
\be
\label{Frn}
\Delta F= - \frac{2\pi^2 r_h L_2^2 }{G_{4} } T^2.
\ee
This  agrees with the  well-known result for a near-extremal RN black hole obtained using the full on-shell action, see eq.(\ref{tfsd}) (with $d=3$).

Note that in the JT model, the  action of the scaling symmetry of $\mathrm{AdS}_2$, under which the  coordinates $(z,t)\rightarrow (\lambda z, \lambda t)$ takes ${\mathcal J}\rightarrow {{\mathcal J}/ \lambda}$ and $T\rightarrow {T/ \lambda}$. To get agreement with the black hole free energy, eq.(\ref{Frn}), we need to fix the scale in the JT model to agree with the  convention for temperature, $T$, at asymptotic infinity.


\section{A More Detailed Comparison with the JT Model}
\label{ss-jtm}

In this section, we explain why  the JT model correctly gives the free energy for  near-extremal black holes. Our explanation is remarkably general; it applies to 
the low-temperature behaviour obtained from  all near-extremal black holes which have a near-horizon $\mathrm{AdS}_2$ region in their geometry.  
We start with considering the Reissner-Nordstr\"{o}m black hole in this subsection, then analyse a more general theory involving scalars, gauge fields and gravity in the following subsection, and finally in subsection \ref{ss-jtrot}, consider rotating black holes. 

One key point in the following discussion is as follows: 
The leading behaviour at low temperature is obtained by computing the on-shell action for a slightly non-extremal solution. This in turn is obtained by keeping the leading corrections in the solution above extremality. Now, away from the near-horizon region, these corrections can be treated as   small perturbations and 
since the extremal black hole  is also a  solution  to the equations of motion,  the contribution due to these  perturbations, at first order, 
 reduces to the  surface terms, eq.(\ref{condda}), at the boundary of the $\mathrm{AdS}_2$ region.  In the near-horizon region, some of the corrections cannot be treated as perturbations, but here their contribution to the bulk action is correctly captured by the JT model action, eq.(\ref{JTact}).

\subsection{RN Black Holes}
\label{ssec-mdrn}
Let us now give more details for the RN case. Carrying out the dimensional reduction described above, we get the two-dimensional action, 
eq.(\ref{dimreda}).  
The resulting radial integral involved in  eq.(\ref{dimreda}) extends from the horizon to asymptotic infinity; we can write this as a sum of two terms: 
\be
\label{fullact}
I=I_{[H\rightarrow \partial \mathrm{AdS}_2]}+I_{[\partial \mathrm{AdS}_2\rightarrow \infty]}.
\ee
Here $I_{[H\rightarrow \partial \mathrm{AdS}_2]}$ denotes the contribution from the horizon to the boundary of asymptotic $\mathrm{AdS}_2$ region described above while  $I_{[\partial \mathrm{AdS}_2\rightarrow \infty]}$ denotes the contribution from the boundary of the asymptotic $\mathrm{AdS}_2$ region to asymptotic infinity. In particular, we take, 
\begin{align}
I_{[\partial \mathrm{AdS}_2\rightarrow \infty]} &= - \frac{1}{4G_{4} } \int_{\del \mathrm{AdS}_2}^\infty \md^2 x \, \sqrt{g} \left(  \Phi^2 R + \frac{2\Phi_0}{\Phi} - 2\Phi_0 \Phi \Lambda_4 + \frac{\Phi^3}{\Phi_0} F^{\alpha \beta} F_{\alpha \beta}  \right) \nonumber \\  
&\quad - \frac{1}{2G_{4} } \int_{\mathcal{\infty}} \md x \, \sqrt{\gamma}\, \left( \Phi^2 K - \frac{2\Phi^3}{\Phi_0} n_\alpha F^{\alpha \beta} A_\beta \right)  + I_{\mathrm{CT}}^\infty  \label{vals2} \\
&\quad - \frac{1}{G_{4} } \int_{\partial \mathrm{AdS}_2} \md x\,\sqrt{\gamma} {\Phi^3\over \Phi_0} n_{\alpha} F^{\alpha \beta} A_\beta . \nonumber 
\end{align}
Here we have included  the boundary terms in eq.(\ref{dimreda}) evaluated at infinity, as well as a counter-term $I^\infty_{CT}$, also to be evaluated at infinity, 
which is  needed to get a finite result, as 
per the standard calculational procedure for computing the free energy.  Finally,  we have included an additional  boundary term at the $\mathrm{AdS}_2$ boundary -- this is the last term above involving the gauge field. In this term, the normal one-form $n_\alpha$ is taken to point from the $\mathrm{AdS}_2$ region towards asymptotic infinity. 
The contribution from the near-horizon region is given by  
\begin{align}
I_{[H\rightarrow \partial \mathrm{AdS}_2]} &= - \frac{1}{4G_{4} } \int^{\del \mathrm{AdS}_2}_H \md^2 x \, \sqrt{g} \left(  \Phi^2 R + \frac{2\Phi_0}{\Phi} - 2\Phi_0 \Phi \Lambda_4 + \frac{\Phi^3}{\Phi_0} F^{\alpha \beta} F_{\alpha \beta}  \right) \nonumber \\
&\quad + \frac{1}{G_{4} } \int_{\partial \mathrm{AdS}_2} \md x\,\sqrt{\gamma} {\Phi^3\over \Phi_0} n_{\alpha} F^{\alpha \beta} A_\beta. \label{vals1}
\end{align}
The bulk integral extends from the horizon to the $\mathrm{AdS}_2$ boundary and the additional  boundary term exactly cancels the last term in eq.(\ref{vals2}).

We will be interested in the contribution made in the on-shell action due to the finite temperature deformation of the solution, at leading order in $T$, keeping the charge $Q$ fixed.
Consider first the near-horizon term, eq.(\ref{vals1}). We denote the contribution  this term makes by $\delta I_{ [H\rightarrow \partial \mathrm{AdS}_2]}$. The 
 dimensionless parameter  characterising the solution is $T/\mathcal{J}$, where $\mathcal{J}$ is determined by the linear variation of the dilaton, eq.(\ref{linvardil}), eq.(\ref{valJ}). Note that 
 ${T/ {\cal J}}\ll 1$ at  low temperatures. The leading contribution in this parameter 
can be obtained  by expanding the dilaton $\Phi$ in terms of the correction $\phi$, eq.(\ref{defphi}),  and only keeping terms which are of first order in $\phi$.
This is because higher order terms will be suppressed by additional powers of $1/\mathcal{J}$ and will therefore be small. This justifies why the last line in eq.(\ref{act3}), involving quadratic and higher powers of $\phi$ can be neglected. 
From the discussion in the previous section, it then follows that 
\be
\label{changea}
\delta I_{ [H\rightarrow \partial \mathrm{AdS}_2]}=\delta I_{\mathrm{JT}}^{\mathrm{ bulk}},
\ee
where 
\be
\label{bulkJT}
I_{\mathrm{JT}}^{\mathrm{ bulk}} = -{1\over {16 \pi \widetilde G}}   \int \md^2 x \, \sqrt{g} \, \phi \left( R-\Lambda_{2} \right),
\ee
with ${\widetilde G}$ given by eq.(\ref{valgt}).

Next consider the finite temperature contribution made from the region away from the $\mathrm{AdS}_2$ horizon, eq.(\ref{vals2}), which we denote by $\delta I_{[\partial \mathrm{AdS}_2 \rightarrow  \infty]}$. A black hole at small temperature is given by the solution eq.(\ref{neRN}) where $f(r)$ takes the form,
 \begin{eqnarray}
 \label{formf}
 f(r) & = & f_{\mathrm{ext}} (r) + \delta f (r),
 \end{eqnarray}
 where
 \begin{align}
 f_{\mathrm{ext}} (r) &= \frac{(r-r_h)^2}{r^2} \left( 1 + \frac{r^2 + 2 r r_h + 3r_h^2}{L^2} \right), \label{deffext} \\
  \delta f (r) &=  -{2 G_{4}  \delta M \over r} \label{defdelf} ,
 \end{align}
 with $\delta M$ being the mass above extremality. In the region being considered here, $f_{\mathrm{ext}}$ does not vanish and for small $\delta M$, $\delta f$ can be treated as a small perturbation, since ${\delta f / f_{\mathrm{ext}} }\ll 1$. As was discussed above, the change in the action then comes from a surface term evaluated at the 
 boundary of $\mathrm{AdS}_2$.

 Some of the steps are described in appendix \ref{app-jttd}. 
 One finds that 
 \be
 \label{condda}
 \delta I_{[\partial \mathrm{AdS}_2\rightarrow \infty]}=
 \delta I_{\mathrm{GH}} + \frac{1}{8\pi \widetilde{G} } \int_{\partial \mathrm{AdS}_2} \md x\, \delta ( \sqrt{\gamma})  \phi K ,
 \ee
 where $I_{\mathrm{GH}}$ is given in eq.(\ref{GHJT}) and both terms  are   to be evaluated at the boundary of the asymptotic $\mathrm{AdS}_2$ region. 
 Also, $\delta ({\sqrt{\gamma}})$ arises from the 
 first order change in the boundary metric. 
 
 The reader will recall from our discussion of the previous section that the counter-term (\ref{jt-ct}) is chosen to cancel the divergence in the GH term, eq.(\ref{GHJT}). 
 As argued in appendix \ref{app-jttd}, it then follows from the  behaviour of the metric in the asymptotic $\mathrm{AdS}_2$ region  that  
 \be
 \label{agreec}
  \frac{1}{8\pi \widetilde{G} }  \int_{\partial \mathrm{AdS}_2} \md x\, \delta ( \sqrt{\gamma})  \phi K =   \frac{1}{8\pi \widetilde{G} L_2}  \int \md x\, {\delta (\sqrt {\gamma}} ) \phi =\delta I_{\mathrm{CT}},
 \ee
 where we have used eq.(\ref{jt-ct}).
 From eq.(\ref{condda}), eq.(\ref{agreec})  we learn  that 
 \be
 \label{relaac}
  \delta I_{[\partial \mathrm{AdS}_2\rightarrow \infty]} =\delta I_{\mathrm{JT}}^{\mathrm{boundary}},
 \ee
 where 
 \be
 \label{jtbdry}
I_{\mathrm{JT}}^{\mathrm{boundary}}=  -\frac{1}{8\pi \widetilde{G} } \int_{\partial \mathrm{AdS}_2} \md x\,  \sqrt{\gamma} \,  \phi \left( K - \frac{1}{L_2}\right).
 \ee
 
 Combining eq.(\ref{changea}) and eq.(\ref{relaac}) we see that the full change in the action is correctly given by the JT model. 
 Note that for  establishing this result, it was important that the boundary conditions, eq.(\ref{bcdil}) and  eq.(\ref{condmet}) are met at the $\mathrm{AdS}_2$ boundary.

 \subsubsection{Some Comments}
 \label{ss-jtsc}

 A few comments are worth making at the end of this subsection. First,
 we note that for the asymptotically flat case, $r_h=Q \sim q'/M_{\mathrm{Pl}}$ (where $q'$ is the dimensionless charge carried by the black hole) and $L_2=r_h$; so that the condition eq.(\ref{valT}) becomes 
 \be
 \label{condTf}
 Tr_h \sim \frac{T q'}{M_{\mathrm{Pl}} } \ll 1.
 \ee
 In the asymptotically $\mathrm{AdS}_4$ case, on the other hand,  for a big black hole with $r_h\gg L$,  $L_2= L/\sqrt{6}$, while  the chemical potential (as seen by the boundary CFT for example) is $\mu ={r_h / L^2} \sim r_h/L_2^2 $.  Thus eq.(\ref{valT})  can be stated as, 
 \be
 \label{condTads}
 T\ll \mu.
 \ee
  
  Second, it is straightforward to generalise the discussion above to RN black holes in arbitrary dimensions. 
  The action for the canonical partition function in this case is given by 
  \be 
  \label{gendac}
I= -\frac{1}{16\pi G_{d+1}} \int \md^{d+1} x\, \sqrt{g} \, \left( R - 2 \Lambda_{d+1} + F_{\mu \nu}  F^{\mu \nu} \right)   - \frac{1}{8\pi G_{d+1}} \int \md^{d} x\, \sqrt{\gamma}\, \left( K - 2 n_\mu F^{\mu \nu} A_\nu \right).
  \ee
  Performing a dimensional reduction by writing the metric as in eq.(\ref{and}), and expanding about the attractor value for $\Phi$ gives a JT model, with ${\widetilde G}$ given in eq.(\ref{geffd}), and ${\cal J}$ given by eq.(\ref{mjd}). As a result, from the JT model, we get the free energy to be eq.(\ref{fJT}) which, in fact, agrees with the result as obtained in eq.(\ref{tfsd}). See appendix \ref{app-gd} for more details.

  Third,  while it is intuitively clear that the low temperature thermodynamics should arise from the near-horizon region, 
  and the well-known diverging back-reaction to perturbations in $\mathrm{AdS}_2$ \cite{Maldacena:1998uz} suggests that at least a linear variation of the dilaton must be incorporated, what the above argument makes precise is that to get agreement with the JT model, it is crucial that the boundary of the near-horizon region is located at a fixed value of the dilaton, eq.(\ref{bcdil}).

  Finally, the agreement with the JT model  for thermodynamics can also be extended for  the response of the black hole to low-frequency   time-dependent probes, as discussed in \cite{Nayak:2018qej,Moitra:2018jqs}\footnote{It is important to include the  extra phase mode and its couplings to charged probes, which we have not described here.}. Higher partial waves can also be systematically incorporated in the two-dimensional theory, and in fact, a precise version of the near-$\mathrm{AdS}_2$ near-$\mathrm{CFT}_1$ correspondence can be then given for these systems \cite{Moitra:2018jqs}. 
  
  \subsection{More General Theories}
  \label{ss-jtgt}
  The arguments above, showing that the thermodynamics at low energies for near-extremal RN black holes is correctly obtained from the JT model,
  can be extended  to more general theories involving additional scalars and gauge fields. 
  Let us  take a theory in $d+1$ dimensions consisting of $m$  gauge fields, $n$ scalars and gravity, with two-derivative interactions given by the action
  \begin{align}
  I &=  - \frac{1}{16\pi G_{d+1}} \int \md^{d+1} x \, \sqrt{g}  \left[ R +  {\hat f}_{ab} (\Psi_i) F_{a \mu\nu}F_b^{\mu\nu} + {\hat h}^{pq}( \Psi_i) g^{\mu \nu} \nabla_\mu \Psi_p \nabla_\nu \Psi_q
  + {\hat V}(\Psi_i) \right] \nonumber \\
  &\quad - \frac{1}{8\pi G_{d+1}} \int \md^{d} x\, \sqrt{\gamma} \, K   \label{actag1}.
  \end{align}

  Here, the indices range from $a,b=1, \cdots, m$, and $ p,q=1, \cdots, n$, ${\hat f}_{ab}$ are gauge kinetic energy coefficients, ${\hat h}_{pq}$ is a metric in the space of scalars and ${\hat V}$ is a potential.  Repeated indices are summed over in eq.(\ref{actag1}).
   In four dimensions, we can also have $F_{\mu\nu} {\widetilde F}^{\mu\nu}$ type of terms and the black holes can also carry magnetic charges. Similarly, in higher dimensions we can add higher form fields under which the black hole carries magnetic charges. We do not explicitly consider these possibilities below although they can be easily included.
  
  To carry out a dimensional reduction, we take the metric to have the form 
  \be
  \label{formmet}
  \md s^2=\left({\Phi_0\over \Phi} \right)^{d-2} g_{\alpha \beta} \,  \md x^\alpha \md x^\beta + \Phi^2 \md\Omega_{d-1}^2,
  \ee
  where $g_{\alpha\beta}$ refers to the metric in the radial and time directions 
    This gives  from eq.(\ref{actag1}) a two-dimensional  action,
  \begin{align}
I=&  - \frac{v_{d-1} }{16 \pi  G_{d+1}} \int \md^{2} x \, \sqrt{g}\left[\Phi^{d-1} R + V( \Psi_i, {\Phi} ) + f^{ab}(\Phi, \Psi_i) F_{a \alpha \beta}F_b^{\alpha \beta}  + h^{pq}(\Phi, \Psi_i) g^{\alpha \beta} 
  \nabla_\alpha \Psi_p \nabla_\beta \Psi_q \right]  \nonumber \\
  & - \frac{v_{d-1} }{8 \pi  G_{d+1}} \int \md x\, \sqrt{\gamma} \, \Phi^{d-1} K  + \frac{v_{d-1} }{4 \pi  G_{d+1}} \int \md x\, \sqrt{\gamma} \, f^{ab}(\Phi, \Psi_i) n_\alpha F_{a}^{ \alpha \beta} A_{b\beta}.\label{actag}
  \end{align}
      One scalar -- the dilaton $\Phi$  is special; it appears in the coefficient of the Ricci scalar, $R$. $\Phi_0$ is its attractor value at the horizon. We have also added an extra boundary term dependent on the gauge field -- the last one above -- since we are interested in the free energy at fixed values of the electric charge. Here,
      $v_{d-1}$ denotes  the volume of the unit $S^{d-1}$, eq.(\ref{dfvd}).

  We are  considering systems which have  an extremal black hole solution  with a near-horizon $\mathrm{AdS}_2\times S^{d-1}$ region.
      The near-horizon region is  well known to be an attractor in these cases.   The gauge fields  in this region take the values
  \be
  \label{valg}
  F_a= \frac{1}{2} f_{ab} Q^b  \sqrt{g} \epsilon_{\alpha \beta} \, \md x^\alpha \wedge \md x^\beta,
  \ee
  where $Q^b$ are the charges carried by the black hole, $f_{ab}$ is the inverse of $f^{ab}$, and $ \frac{1}{2}\sqrt{g} \epsilon_{\alpha \beta} \, \md x^\alpha \wedge \md x^\beta$ is the volume form in $\mathrm{AdS}_2$.  The scalars in the near-horizon region  take the attractor values  $\{\Phi_0,\Psi_{0i}\}$.
  The resulting  attractor values for the scalars can be computed by extremising the entropy function  \cite{Sen:2005wa, Sen:2007qy} 
   or by using an effective potential \cite{Goldstein:2005hq}.

   The free energy is given by the on-shell value of this action evaluated for a black hole at temperature $T$ carrying charges $Q^a$. 
   The action in eq.(\ref{actag})  is again a sum of two terms. One, involving the integral from the horizon to the asymptotic $\mathrm{AdS}_2$ region and the second, from the asymptotic $\mathrm{AdS}_2$ region to asymptotic infinity. 
   We denote them, as above, by $I_{[H\rightarrow \partial \mathrm{AdS}_2]}$ and $I_{[\partial \mathrm{AdS}_2 \rightarrow \infty]}$ respectively, these are analogous to 
   eq.(\ref{vals1}) and eq.(\ref{vals2}) above.
   
 In the near-horizon region of the extremal solution, once one is away from the horizon, the dilaton evolves with a linear radial dependence, e.g., eq.(\ref{pdv}).
 Similarly, the other scalars $\Psi_i$  also evolve along the radial direction and their evolution gives rise to scales $\mathcal{J}_i$ which are analogous to the scale $\mathcal{J}$ that characterises the breaking of scale invariance due to the non-constant dilaton. We will take all these scales to be comparable to simplify the discussion, i.e., $\mathcal{J} \sim \mathcal{J}_i$. 
 The condition for temperature $T$ to be low enough is then given by 
 \be
 \label{condgt}
 T\ll \mathcal{J}, \, \mathcal{J}_i.
 \ee
  Expanding the scalars about their attractor values, 
  \begin{align}
  \Phi &= \Phi_0(1+\phi), \nonumber \\
  \Psi_i &= \Psi_{0i} + \delta \Psi_i.   \label{scexp}
  \end{align}
   we locate the boundary of  the $\mathrm{AdS}_2$ region at a location where the dilaton takes a fixed value, $\phi_B$, with the condition,  eq.(\ref{valzb}),
      being met.
    The other scalars $\delta \Psi_i$  do not take fixed values at this boundary, but it follows from eq.(\ref{valzb}), eq.(\ref{condgt})
   that their boundary values meet the condition, $\delta \Psi_{Bi}/\Psi_{0i} \ll 1$. 
  
  In the near-horizon region,  for calculating the contribution $I_{[H\rightarrow \partial \mathrm{AdS}_2]}$ makes to the free energy, we can work to linear order in $\phi$ and $\delta \Psi$, since higher order terms will be suppressed by higher powers of $\mathcal{J}$, $\mathcal{J}_i$ and will thus be suppressed at low $T/\mathcal{J}$. 
  It is then easy to see that the contribution of the near-horizon region is given by eq.(\ref{bulkJT}) with ${\widetilde G}$ taking the value, 
  \be
   \label{valgtg}
   {\widetilde G}= \frac{ G_{d+1}}{v_{d-1}(d-1) \Phi_0^{d-1}}   . \ee
  
  In the region away from the horizon, on the other hand, the leading order changes in the metric etc., can all be treated as small perturbations, as in the RN case. 
  Thus the resulting contribution to $\delta I_{[\partial \mathrm{AdS}_2 \rightarrow \infty]}$ comes from  surface terms at the $\mathrm{AdS}_2$ boundary. Surface terms which arise from the curvature dependent term, $\Phi^{d-1} R$, in eq.(\ref{actag}) give rise, at linear order in $\phi$ to the boundary terms in  the JT model, eq.(\ref{jtbdry}), this follows from an argument analogous to the RN case discussed in appendix \ref{app-jttd}. 
  Surface terms arising from the scalar kinetic energy terms in eq.(\ref{actag}) do not lead to any contribution at leading order, as explained in appendix \ref{app-jttd}. 
  As a result, once again, we find that the JT model correctly gives the leading free energy at small $T$.

   Let us end with some comments.   It is easy to argue that the leading departure away from the attractor value for $\phi$ takes the form eq.(\ref{linvardil}), in fact this follows from the equations of motion in the JT model. 
   $\mathcal{J}$ is now an energy  scale that depends in general  on all  the charges $Q^a$, as well as the asymptotic data at infinity.  Also, the resulting free energy  is given by eq.(\ref{fJT}) with ${\widetilde G}$ being given by eq.(\ref{valgtg}). 
   Note that as was  mentioned above, the other scalars also evolve and this gives rise to additional sources for the breaking of scale invariance; however quite notably, it is the breaking due to the evolution of the dilaton that  determines the free energy.

\subsection{Rotating Black Holes}
\label{ss-jtrot}
The metric for a rotating black hole is well known to be quite complicated, as will also be apparent below. 
Despite these complications, quite interestingly, the  discussion above can be extended to such black holes as well in a fairly straightforward way.

A few important features come to our aid in simplifying the analysis and connecting it to the JT model. 

First, many rotating extremal black holes are also known to posses an $\mathrm{AdS}_2$ symmetry in their near-horizon geometry and our 
analysis applies  only to such cases. In fact, this property is true   quite generically for rotating black holes when their extremal entropy is non-vanishing. 
For example, for  an extremal $4$-dimensional Kerr black hole \cite{Kerr:1963ud}, in asymptotically flat spacetime, the near-horizon geometry 
is  \cite{Bardeen:1999px}
\begin{equation}
\label{4dflkatment}
\md s^2 = \frac{1 + \cos^2 \theta}{2} \Big(  \frac{r^2}{2{r_h}^2} \md t^2 + \frac{2 {r_h}^2}{r^2} \md r^2 +   2 {r_h}^2 \, \md  \theta^2 \Big) + 
\frac{4{r_h}^2 \sin^2 \theta}{(1+ \cos^2 \theta)} \Big( \md \varphi -\mi  \frac{r }{2{r_h}^2}  \md t\Big)^2.
\end{equation}
In these coordinates, the horizon is located at $r=0$ (the radius of the event horizon is $r_h$ in Boyer-Lindquist coordinates \cite{Boyer:1966qh}; see below). This metric  exhibits the  $\mathrm{AdS}_2$ symmetries with the Killing vectors being
 \begin{align}
\zeta_{1} &= - L_2 \frac{\partial}{\partial t}, \nonumber \\
\zeta_0 &= t \frac{\partial}{\partial t} - r \frac{\partial}{\partial r}, \label{sl2r} \\
\zeta_{-1} &=  \frac{L_2}{2} \left( \frac{L_2^2}{r^2} - \frac{t^2}{L_2^2} \right) 
 \frac{\partial}{\partial t} + \frac{t r}{L_2} \frac{\partial}{\partial r} +\mi \frac{L_2^3 \alpha}{r } \frac{\partial}{\partial \varphi}. \nonumber
\end{align}
These generate an $\mathrm{SL}(2, \mathbb{R})$ algebra.  In this case, $L_2^2 = 2r_h^2$ and $\alpha = 1/2r_h^2$.

Second, it has also been argued that rotating  extremal black holes with such an  $\mathrm{AdS}_2$ symmetry exhibit the attractor phenomenon which fixes their near-horizon geometry. The values of various scalars as well as the angular dependence in  the metric are fixed by extremising the entropy function \cite{Astefanesei:2006dd}. 
 
 Finally, despite the fact that for the rotating case rotational invariance is broken and many more components of the metric are  excited, 
 a key feature  that allows the  connection with the JT model to be made for purposes of obtaining the free energy  is the following.  In calculating the leading order temperature dependence of the free energy,  many alterations in the metric, in going from the extremal to slightly non-extremal case,  can be regarded as small perturbations in the whole  of spacetime extending from the horizon to asymptotic infinity. 
 It follows then along the lines of the argument given above in the rotationally invariant case, that  there is no change in the on-shell action at first order due to  these alterations,    since the extremal black hole we perturb about is already a solution of the equations of motion.  Furthermore, it is sufficient to work to first order in these alterations   for obtaining the leading temperature dependence. Therefore these changes in  the metric, while going from the extremal  to the slightly non-extremal case, can be neglected.  
 
 This observation leaves  only a  few remaining changes in the metric which we have to keep track of.  Even for these,  there is a further simplification. 
 Consider computing the free energy by dividing the contribution to the on-shell action as coming from the region close to the horizon, and sufficiently far away from it, analogous to the two terms, eq.(\ref{fullact}),  above. Now even though some changes are not small in the near-horizon region, they are  small perturbations in the region far away from the horizon. As a result, the contribution of this far region can be expressed as a surface term at the boundary of the near-horizon region.  This finally only leaves the near-horizon region, but here,  the $\mathrm{AdS}_2$ symmetry and the attractor behaviour mentioned above come to our aid and allow us to conclude that the JT model is correct!  
 
 Let us give some more details now. For concreteness, we consider the asymptotically flat $4$-dimensional case considered above. 
 The  metric, for a general non-extremal Kerr black hole is 
 \be
\label{metextf1}
\md s^2={\rho^2\Delta_r \over \Sigma} \md t^2 + {\rho^2\over \Delta_r} \md r^2 +  \rho^2 \md\theta^2 +  \sin^2\theta {\Sigma\over \rho^2}(\md\varphi + \mi \, \omega \, \md t)^2,
\ee
where
\begin{align}
\Delta_r &= r^2 - 2 G_4 M r + a^2, \label{flatDelta} \\
\rho^2 &= r^2 + a^2 \cos^2 \theta, \label{flatrho} \\
\Sigma &= (r^2 +a^2)^2 - a^2 \Delta_r \sin^2 \theta , \label{flatSigma} \\
\omega &= \frac{2G_4 M a r}{\Sigma}. \label{flatomega}
\end{align}
The various functions appearing above are obtained by taking the $L\to \infty, q_e \to 0, q_m \to 0$ limit of the functions in eqs.(\ref{xi4}-\ref{om4}). 
 At extremality, $\Delta_r$ has a second order zero at $r=r_h=a_0$. (We have denoted the extremal value of $a$ by $a_0$.) This gives the near-horizon metric in eq.(\ref{4dflkatment}) (with  $\varphi \rightarrow \varphi - \mi \Omega_H t  $, where $\Omega_H = \omega(r_h)$; see also appendix \ref{app-rot4d}).

 The metric, eq.(\ref{metextf1}) depends on  $M$,  the mass and $J$,  the angular momentum. They are related through the parameter $a$, 
 \be
 \label{vala}
 J=M a. 
 \ee
 The extremal mass is given by,
 \be
 M_0 = \frac{r_h}{G_{4}} = \frac{a_0}{G_{4}}.
 \ee
 
 When we go to the slightly non-extremal case keeping $J$ fixed, $M$ changes and from eq.(\ref{vala}) $a$ changes as well. 
 This results in all the functions which appear in eq.(\ref{metextf1})  being altered. However, for some of them the fractional change is small everywhere in spacetime outside the horizon.
For instance, for  $\rho^2$ we have from eq.(\ref{flatrho}) that 
 \be
 \label{chrh}
 \delta\rho^2=2   a_0 \cos^2\theta \, \delta a,
 \ee
 so that ${\delta \rho^2\over \rho^2} \ll 1$  everywhere. 
 
In contrast, the change in $\Delta_r$ while being small in the region from the $\mathrm{AdS}_2$ boundary to infinity, is not small in the near-horizon region, since it vanishes
at $r=r_h$ in the extremal case. Therefore, $\Delta_r$  is a function whose change we need to keep track of in going to the non-extremal case. 

To proceed, let us write the  metric in the form,
\be
\label{metextf}
\md s^2={\rho^2\Delta_r \over \Sigma} \md t^2 + {\rho^2\over \Delta_r} \md r^2 + \Phi^2 \pqty{ {\rho^2 \over \sqrt{\Sigma}}} \md \theta^2 + \Phi^2 \sin^2\theta  \pqty{ {\sqrt{\Sigma} \over \rho^2} }(\md\varphi -\mi  A_t\, \md t)^2.
\ee
Here, 
\be
\label{defphirot}
\Phi^2= \sqrt{\Sigma} = ({(r^2+a^2)^2-a^2 \Delta_r \sin^2 \theta})^{1/2},
\ee
\be
\label{valat}
A_t= \Omega_H - \omega,
\ee
and we are working   in the shifted azimuthal coordinate $\varphi\to \varphi -\mi \Omega_H t$. One advantage of parametrising the metric in this manner is that 
the volume of the internal two sphere spanned by $\theta, \varphi$ is now  given by $4 \pi \Phi^2$ and therefore only dependent on $\Phi$, and manifestly independent of $\Sigma$. Although $\Phi$ is evidently a function of $\Sigma$, as defined above, when $\Phi$ and $\Sigma$ are regarded formally as independent quantities, the volume of the transverse space depends only on $\Phi$. We emphasise the role of $\Phi$ to make the connections with the previous discussion more transparent.
As a result, after dimensional reduction to two dimensions, the dilaton field will be proportional to $\Phi^2$ (as in eq.(\ref{dimredb})), and the variation of $\Phi$ from its attractor value (\ref{defphi}) will be important 
 in obtaining the thermodynamics. In contrast, by repeating the analysis above for $\rho^2$, now for the field $\Sigma$, it is easy to see that its change from the extremal value will not be important in obtaining  the leading order change in the free energy.   We also note that we will need to keep track of the change in $\Delta_r,$ and $A_t$, when going to the non-extremal case, for computing the corrections to the free energy.

It is now easy to obtain the bulk action in the near-horizon region. The metric in this region can be written as 
\be
\label{metformaa}
\md s^2={1+\cos^2\theta\over 2} g_{\alpha\beta} \, \md x^\alpha \, \md x^\beta + {1+\cos^2\theta\over 2} \Phi^2 \,  \md \theta^2 + \Phi^2 
{2 \sin^2\theta\over 1+ \cos^2\theta} (\md \varphi- \mi A_\alpha \md x^\alpha)^2.
\ee
Here, $\alpha, \beta$ take values in $r,t$.
 
Starting from the action
\be
\label{acta}
I=-{1\over 16 \pi G_{4} }\int \md^4 x \sqrt{g} R,
\ee
and inputting the metric eq.(\ref{metformaa}), we obtain,
\begin{align}
\label{twodimmeta}
I&=- {1\over 4 G_{4} }\int \md^2x \sqrt{g} \left( \Phi^2 R + 2 ( \nabla \Phi)^2 + 1 + \frac{1}{2}\Phi^4 F_{\alpha \beta} F^{\alpha \beta} \right) + \frac{1}{G_{4} } \int_{\partial}\md x \sqrt{\gamma} n^\alpha \Phi \nabla_\alpha \Phi ,
\end{align}
where $R$ which appears in eq.(\ref{twodimmeta}) is the Ricci curvature of the two-dimensional metric. This reduction, a special case of the reduction we have carried out in section \ref{sec-4drbf}, (zero electric and magnetic charges and zero cosmological constant limit) was also considered in \cite{Castro:2009jf}.
 
We are interested in a rotating black hole of angular momentum $J$; the field strength in this case is given by 
 \be
 \label{fst}
 F_{\alpha\beta}=\frac{2J G_{4} }{\Phi^4} \sqrt{g} \epsilon_{\alpha \beta} .
 \ee
 
 Expanding $\Phi$ as given in eq.(\ref{defphi}), where 
 \be
 \label{defphi0}
 \Phi_0= \sqrt{2 G_{4}  J},
 \ee
 and carrying out a change in frame
 \be
 \label{framechng}
 g_{\alpha\beta}\rightarrow \frac{\Phi_0}{\Phi} g_{\alpha\beta} 
 \ee
 gives, 
 for the action eq.(\ref{twodimmeta}),
 \be
 \label{actra}
 I=- \frac{\Phi_0^2}{2 G_{4 } }\int \md^2 x \, \sqrt{g} \, \phi(R-\Lambda_{2} ) + \frac{3\Phi_0^2}{4 G_{4} }\int_{\partial} \md x \, \sqrt{\gamma} n^\alpha \nabla_\alpha \phi - \frac{\Phi_0^4}{2G_{4}} \int_\partial \md x \, \sqrt{\gamma} (1+ 5\phi )n_\alpha  F^{\alpha \beta} A_\beta.
 \ee
 Here we have only included terms  to linear order in $\phi$ and not included a topological term  leading to  the ground state entropy. We note that the bulk term above agrees with the JT model.

We had mentioned that the region between the $\mathrm{AdS}_2$ boundary and infinity contributes a surface term at the $\mathrm{AdS}_2$ boundary. We add to this the two surface terms in eq.(\ref{actra}). A convenient way of calculating the resulting total of all surface terms  is the following. 
We have argued that the changes in the two-dimensional metric etc.\ that we are interested in cannot be regarded  as perturbations in the $\mathrm{AdS}_2$ region.
But we could consider other changes which would be perturbations in the whole of spacetime outside the horizon. 
For such changes, the  total alteration in the action to first order in these perturbations from the near $\mathrm{AdS}_2$ region should also have been a 
surface term at the boundary of $\mathrm{AdS}_2$ which cancels the surface term which arises from the region between the $\mathrm{AdS}_2$ boundary and infinity. 
This along with general covariance in the two-dimensional theory allows us to obtain the surface terms. 

Let us denote by $I_{[H\rightarrow \partial \mathrm{AdS}_2]}$ the bulk term in eq.(\ref{actra}):
\be
\label{bulkactra}
I_{[H\rightarrow \partial \mathrm{AdS}_2]}=- \frac{\Phi_0^2}{2 G_{4} }\int \md^2 x \, \sqrt{g} \, \phi(R-\Lambda_2).
\ee
The change in the $\mathrm{AdS}_2$ region bulk action for perturbations  which satisfies the  Dirichlet boundary condition for $\phi$ gives, (see appendix \ref{app-jttd}) 
\be
\label{changeact}
\delta I_{[H\rightarrow \partial \mathrm{AdS}_2]} =  {\Phi_0^2\over  G_4}  \int_\partial \md x\, \delta ( \sqrt{\gamma} K) \phi - {\Phi_0^2 \over G_4}  \int_\partial \md x \, \delta (\sqrt{\gamma}) \phi  K.
\ee
Here the extrinsic curvature is computed using a normal which is directed outward towards asymptotic infinity. As a result, the surface term arising from the bulk integral from the $\mathrm{AdS}_2$ boundary to  infinity. which cancels this contribution is opposite in sign from that in eq.(\ref{changeact}). We see that with this change in sign, this contribution then agrees with what is obtained from the boundary terms in the JT model, we have also used the argument in subsection \ref{ssec-mdrn}, eq.(\ref{agreec}) that the second term above equals the contribution from the counter-term. 
 
Since the bulk term in eq.(\ref{actra}) also agrees with the JT model, it then follows that the change in the free energy will agree with the result in the JT model. 
Let us also note that the connection to  the JT model is tied  to the near horizon geometry exhibiting an $\mathrm{SL}(2,\mathbb{R})$ symmetry. This symmetry results in the metric in the $t,r$ plane being conformally $\mathrm{AdS}_2$, eq.(\ref{4dflkatment}), and allows the free energy calculation, after carrying out the integral in the $\theta,\varphi$ directions, to be related to that in the JT model.

The discussion above can be extended to other rotating black holes, including those with extra gauge fields and scalars as well. The crucial point is that the bulk action in the near-horizon region becomes  that of the JT model. It then follows that  the additional surface terms can be also obtained from the near-horizon region and agree with those in the JT model, eq.(\ref{jtbdry}),  as  was illustrated in the Kerr black hole above.

Comparing with the $4$-dimensional Kerr case, it is clear that for a general $d+1$-dimensional spacetime  where we start with the action, 
\be
\label{genacta}
I=-{1\over 16 \pi G_{d+1}} \int \md^{d+1}x \, \sqrt{g} R + \cdots,
\ee
${\widetilde G}$, eq.(\ref{JTact}), in the JT model which arises,  is given by eq.(\ref{valgtg}).
Note that  the volume of the transverse space spanned by the angular directions is 
\be
\label{defV}
V_{d-1}=v_{d-1} \Phi^{d-1},
\ee
with $v_{d-1}$ being the volume of the unit $S^{d-1}$, eq.(\ref{dfvd}), and $\Phi_0$  appearing in eq.(\ref{valgtg}) is the attractor value for $\Phi$. 
As in the cases above, the role of the dilaton in the JT model is played by the linear correction to $\Phi$ when it is expanded about $\Phi_0$ as given in eq.(\ref{defphi}). 
And the scale $\mathcal{J}$, eq.(\ref{linvardil}), is  determined by the linear variation in $\phi$ away from the attractor value. 

We end with some comments. First, for a Kerr-Newman type of black hole, one can consider starting with no angular momentum and then adding some rotation. 
For small angular momentum, it is easy to see that extra Kaluza-Klein (KK) modes, corresponding to higher partial waves along the transverse sphere, are turned on.
In such  cases a two-dimensional reduction can be carried out in the region between the $\mathrm{AdS}_2$ boundary and infinity as well, retaining these extra modes, and the analysis for these cases can be  mapped to that in \S\ref{ss-jtgt}. Second, our arguments for rotating black holes are somewhat less straightforward than for the spherically symmetric ones in the previous subsections. We will consider
several examples below and see that the JT model does correctly give the free energy for the rotating cases as well. This adds  additional evidence in support of these arguments. 

Finally, as a summary of this section we note that  we have seen that the dilaton, in all cases including rotating ones, corresponds to the volume of the transverse sphere spanned by the angular directions, with $\Phi_0$ being related to the area of the horizon, by eq.(\ref{defV}), 
\be
\label{ah}
A_H= v_{d-1} \Phi_0^{d-1}.
\ee
Using eq.(\ref{valgtg}) and eq.(\ref{fJT}) this allows us to express the free energy for near-extremal black holes in $d+1$ spacetime dimensions, as 
\be
\label{finalfree}
\Delta F=-(d-1) \pi S_0 {T^2 \over {\cal J}},
\ee
where $S_0={A_H / 4 G_{d+1}}$ is the ground state entropy.


 \section{Five-Dimensional Rotating Black Holes in Asymptotically Anti-de Sitter Spacetime}
 \label{sec5drf}
  
 We are now ready to consider some specific examples of rotating near-extremal black holes. The first case we consider in this section is that of a five-dimensional rotating black hole in asymptotically $\mathrm{AdS}_5$ spacetime. 
 This example was studied in \cite{Castro:2018ffi} where the connection with the JT model was also explored. We will see that this example can, in fact, be related to the discussion in \S\ref{ss-jtgt}.
  
 We start with the Euclidean action,
 \begin{equation}
\label{n5deinac}
I = -\frac{1}{16\pi G_{5} }\int \md^5 x\, \sqrt{g } \left( R - 2 \Lambda_{5} \right) -\frac{1}{8\pi G_{5} } \int \md^4 x \, \sqrt{\gamma } K.
\end{equation} 
 The metric of the five-dimensional rotating black hole is given by, (see \cite{Castro:2018ffi} and \cite{Hawking:1998kw,Gibbons:2004uw,Gibbons:2004js}) 
 \begin{equation}
\label{5met}
\md s^2 = \frac{\Delta}{\Xi} \exp(U_2 - U_1) \md t^2 + \frac{r^2}{(r^2+a^2)\Delta}\md r^2 + \exp(-U_1) \, \md \Omega_2^2 + \exp(-U_2) (\md\psi+A)^2,
\end{equation}
where $\Delta, U_1,U_2$ are functions of $r$ and $A$ is a  one-form (see appendix \ref{app-rot5d}). Note that the angular coordinate $\psi$ lies in the range $[0,4\pi)$.

For the extremal black hole  the parameter $\Delta$ has a second order zero in $r$. The metric in the near-horizon limit takes the   attractor value 
 \be 
  \label{5attg}
\md s^2 = \frac{1}{2 \left(1 + \frac{r_h^2}{L^2} \right) } \left( \frac{(r-r_h)^2 }{L_2^2} \md t^2 + \frac{L_2^2}{(r-r_h)^2 } \dd r^2  \right) +  \frac{r_h^2}{2  \left(1 - \frac{2r_h^2}{L^2} \right)  } \md \Omega_{2}^2 + \frac{r_h^2}{\left(1 - \frac{2r_h^2}{L^2} \right)^2  } ( \md{\psi} + \bar{A} )^2  ,
  \ee
  where,
\begin{align}
\bar{A} =  \cos \theta \, \md \varphi - \mathrm{i} {\left(1 - \frac{2r_h^2}{L^2} \right) \left(1 + \frac{2r_h^2}{L^2} \right)^{1/2}}  \frac{r-r_h}{r_h^2} \frac{1}{\left(1 + \frac{r_h^2}{L^2} \right)}  \md t \label{5nhg}.
\end{align}
Here $L$ is the $\mathrm{AdS}_5$ radius, (\ref{5dccdef}), and we have transformed the angular coordinate, $\md{\psi} \rightarrow \md \psi +  k \, \md t $, for a constant $k$. We see that in the $r-t$ plane,  at constant values for the angular coordinates, it clearly exhibits an $\mathrm{AdS}_2$ geometry. The radius of $\mathrm{AdS}_2$ is given by,
\be
\label{valads25}
L_2^2 = r_h^2  \frac{\left(1 + \frac{r_h^2}{L^2} \right)}{\left(1 + \frac{4r_h^2}{L^2} \right)}.
\ee

Going to the slightly non-extremal case, keeping the  angular momentum $J$, eq.(\ref{j5}) fixed, it is easy to calculate the entropy and the free energy as a function of temperature $T$.   One finds that 
\be
\label{nhf}
\Delta F \equiv F-F_{\mathrm{extremal}} =      -\frac{\pi}{4 \widetilde G } {T^2 \over \mathcal{J}}.
\ee
We have given the result in terms of  the parameters ${\widetilde G}, \mathcal{J}$ in view of the discussion that follows. These parameters take the values,
\be
\label{valtg5}
{\widetilde G}=    \frac{G_{5} \left( 1 - \frac{2r_h^2}{L^2} \right)^2}{24 \pi^{2} r_h^{3}}, 
\ee
\be
\label{valj5}
\mathcal{J}=\frac{3}{r_h} \frac{1 + \frac{4r_h^2}{L^2}}{1 - \frac{r_h^2}{L^2}}.
\ee

We can now compare this result with the JT model that we obtain for this black hole. 
It is useful to carry out the dimensional reduction in obtaining the  two-dimensional theory  here in two steps. (Note that a similar dimensional reduction was carried out in \cite{Castro:2018ffi}).
First, writing the metric as 
\be
\label{met4}
\md s_{(5)}^2=g^{(4)}_{\mu \nu} \md x^\mu \md x^\nu + \Sigma^2( \md \psi+ A)^2,
\ee
we carry out a dimensional reduction of the standard KK type over the $\psi $ direction. This gives rise to a $4$-dimensional theory, consisting of the metric, $g_{\alpha\beta}$, a gauge field $A$, and the scalar $\Sigma$. This theory is of the kind considered in \S\ref{ss-jtgt}. The  action of the $4$-dimensional theory is 
\be
\label{actfd}
I= - \frac{1}{4 G_{5} } \int \md^4 x \sqrt{g^{(4)} } \left( \Sigma R -  \frac{1}{4} \Sigma^3 F_{\mu \nu} F^{\mu \nu} - 2 \Sigma \Lambda_{5} \right) -  \frac{1}{2 G_{5} } \int \md^3 x \sqrt{\gamma^{(3)} }  \Sigma K 
\ee
(which  is closely related to the  type of action considered in \S\ref{ss-jtgt}). The gauge field here carries one unit of magnetic charge and also electric charge which is related to the angular momentum carried by the five-dimensional black hole. 
It is clear from eq.(\ref{5met}) that the $4$-dimensional metric obtained after KK reduction is rotationally symmetric.

To come down to two dimensions, we next take the four-dimensional metric to have the form
\be
\label{fourdmet}
\md s_{(4)}^2=\frac{1}{\Sigma \Phi^{3/2}} g_{\alpha\beta} \md x^\alpha \md x^\beta + \frac{\Phi^3}{8\Sigma} \md\Omega_{2}^2.
\ee
The indices $\alpha,\beta$ refer to the two-dimensional coordinates. Here $\Phi^3$ is related to the $3$-volume $V_3$  spanned by the  $\theta,\varphi,\psi$ directions as follows:
\be
\label{defphia}
V_3=v_3\Phi^3,
\ee
where $v_3$ is the volume of a unit $S^3$. 

Inserting eq.(\ref{fourdmet}) in  eq.(\ref{actfd}), with the gauge field ansatz,
\begin{equation}
\label{gfa5}
A_\mu \md x^\mu = \cos \theta \, \md \varphi + \mathrm{i} A_\alpha \md x^\alpha,
\end{equation}
gives the two-dimensional action, eq.(\ref{kads5-2dac})  of appendix \ref{app-rot5d}. This action is of the form given in eq.(\ref{actag}).

 From eq.(\ref{5met}) we see that that 
\be
\label{vols35}
\Phi^3 = 8 \exp(-U_1 - \frac{1}{2} U_2).
\ee
Its attractor value is given by,
\be
\label{attracphi}
\Phi_0^3=   \frac{4r_h^3}{\left( 1 - \frac{2r_h^2}{L^2} \right)^2}.
\ee

Expanding $\Phi$ about the attractor value, eq.(\ref{defphi}), we get that $\phi$ varies linearly as given in eq.(\ref{linvardil}) with $z$ being related to the radial coordinate $r$  in eq.(\ref{coordt}), and $\mathcal{J}$ taking the value,  eq.(\ref{valj5}). 
From eq.(\ref{a5jt}) and eq.(\ref{JTact}) we see that  that ${\widetilde G}$ takes the value, eq.(\ref{valtg5}).
 
It then follows from eq.(\ref{fJT})  that the free energy of the JT model is also given by eq.(\ref{nhf}) and agrees with the result above. More details are given in appendix \ref{app-rot5d}. 

Let us note that in \cite{Castro:2018ffi}, the authors  in their analysis of the JT model found an ambiguity in various thermodynamic parameters which corresponds to a choice of scale in defining the temperature. We have  chosen   conventions  here so that the  temperature agrees with the definition at asymptotic infinity.

We end this section with two comments. First, the free energy eq.(\ref{nhf}) depends on $r_h$ which in turn can be obtained in terms of the angular momentum from 
eq.(\ref{j5}), eq.(\ref{kads5-mext}), eq.(\ref{kads5-aext}). Second these black holes, for $L\ne \infty$ could suffer from a superradiant instability, as discussed in \cite{Hawking:1999dp,Kunduri:2006qa}, see also appendix \ref{app-rot5d}. The flat spacetime limit $L\rightarrow \infty$ does not have this instability.


\section{Four-Dimensional Charged Rotating Black Holes}
\label{sec-4drbf}
Next we study four-dimensional rotating black holes. With the  strength we have by now  acquired,  we might as well consider the most general case of a Kerr-Newman near-extremal black hole  with both electric and magnetic charges in asymptotically $\mathrm{AdS}$ spacetime! The uncharged  near-extremal Kerr black hole in flat spacetime is a special case which we will comment on in more detail later on in the section.
 
 The Euclidean action is,
 \begin{equation}
\label{dac4}
I = -\frac{1}{16\pi G_{4} } \left[ \int \md^4 x\, \sqrt{g } \left( R - 2 \Lambda_{4} - \hat{F}_{\mu \nu} \hat{F}^{\mu \nu} \right) + 2 \int \md^3 x \, \sqrt{\gamma } \, K \right].
\end{equation}

There is a well-known rotating black hole solution given by \cite{Carter:1968ks,Plebanski:1976gy,Hartman:2008pb}.
\begin{equation}
\label{mt4}
\md s^2 =  \frac{\rho^2 \Delta_r \Delta_\theta}{\Sigma} \md t^2 + \frac{\rho^2}{\Delta_r} \md r^2 + \frac{\rho^2}{\Delta_\theta} \md \theta^2 + \frac{\Sigma}{\rho^2 \Xi^2} \sin^2 \theta ( \md \varphi + \mathrm{i} \, \omega \, \md t)^2.
\end{equation}
where,
\begin{align}
L  &= \sqrt{-\frac{3}{\Lambda_{4} } }, \label{la4}\\
\Xi &= 1 - \frac{a^2}{L^2}, \label{xi4} \\
\rho^2 &= r^2 + a^2 \cos^2 \theta, \label{ro4} \\
\Delta_r &= (r^2 + a^2 ) \left(  1 + \frac{r^2}{L^2} \right) - 2   m r + (q_e^2 + q_m^2),  \label{dr4}\\
\Delta_\theta &= 1 - \frac{a^2}{L^2} \cos^2 \theta, \label{dt4} \\
\Sigma &= (r^2 + a^2)^2 \Delta_\theta - a^2 \Delta_r \sin^2 \theta, \label{s4} \\
\omega &= \frac{a \Xi [ (a^2 + r^2) \Delta_\theta -\Delta_r]}{\Sigma}. \label{om4}
\end{align}

The gauge field is given by,
\begin{equation}
\label{ggf4}
\hat{A} =  \frac{q_e r}{\rho^2} \left( \mi \, \md t + \frac{a \sin^2 \theta}{\Xi} \md \varphi \right) + \frac{q_m \cos \theta (r^2 + a^2)}{\Xi \rho^2} \left(  \md \varphi + \mathrm{i}  \frac{a \Xi}{r^2 + a^2}  \, \md t \right).
\end{equation}

The ADM mass, the angular momentum, the physical electric and magnetic charge are given by,
\begin{align}
M &= \frac{m}{G_{4} \Xi^2}, \label{madm4} \\
J &= \frac{ma}{G_{4} \Xi^2}, \label{kj4} \\
Q_e &= \frac{q_e}{\Xi}, \label{qe4} \\
Q_m &= \frac{q_m}{\Xi}, \label{m4}
\end{align}
respectively.

Note that as in the five-dimensional case, (see appendix \ref{app-rot5d}) $|a|$ can approach $L$ only from below.

As discussed in appendix \ref{app-rot4d}, close to extremality, the change in entropy, above its extremal value, $\Delta S$ is given by 
\be
\label{4rots}
\Delta S= { \pi \over {2 \widetilde G} } {T\over {\cal J}},
\ee
where we have defined
\begin{equation}
\label{knag}
\widetilde{G} = G_{4} \frac{\left( 1 - \frac{a_0^2}{L^2} \right)}{8\pi (r_h^2 + a_0^2) },
\end{equation}
and 
\begin{equation}
\label{knaj}
\mathcal{J} = \frac{1}{r_h }  \left( 1 + \frac{6r_h^2 + a_0^2}{L^2} \right), 
\end{equation}
with an eye towards the discussion of the JT model that follows. Here, $r_h$ is the radial location of the extremal horizon and quantities with a subscript 0 denote the values of these parameters at extremality. For more details, see appendix \ref{app-rot4d}.
Using thermodynamic relations, we then get the free energy above extremality to be,
\be
\label{kfree}
\Delta F=-{ \pi \over {4 \widetilde G} } {T^2\over {\cal J}}.
\ee
This result can also be obtained directly by calculating the on-shell action. 

Let us now connect this discussion to the JT model. Our discussion will follow  \S\ref{ss-jtrot}.
The near-horizon metric for the extremal case is given by 
\begin{align}
\md s^2 &= \frac{\rho_0^2 }{(r_h^2 + a_0^2) } \left(  \frac{(r-r_h)^2}{L_2^2} \md t^2 + \frac{L_2^2}{(r-r_h)^2} \md r^2 \right)  + \frac{\rho_0^2 \Xi_0 }{(r_h^2 +a_0^2) \Delta_{\theta 0} } \Phi_0^2 \, \md \theta^2 \nonumber \\
&\quad + \frac{\Delta_{\theta 0} (r_h^2 + a_0^2)  }{\Xi_0 \rho_0^2} \Phi_0^2 \sin^2 \theta \left( \md {\varphi} - \mathrm{i} \frac{2  r_h a_0 \Xi_0}{(r_h^2 +a_0^2)^2 } (r-r_h) \, \md t \right)^2, \label{nhrota}
\end{align}
For further details, including the definitions of the various quantities appearing above, see appendix \ref{app-rot4d}. For the uncharged, asymptotically flat case, with $q_e,q_m=0, L \to \infty$, this agrees with eq.(\ref{4dflkatment}).
The  gauge fields  in the near-horizon region takes the form given below  in eq.(\ref{gfan4}) with
\be
\mathcal{A}_t =  \frac{q_{e0} (r_h^2 - a_0^2)}{(r_h^2 + a_0^2)^2} (r-r_h) , \quad \mathcal{A}_r  = 0, \label{gf4r}
\ee
\be
A_t = \frac{2  r_h a_0 \Xi_0}{(r_h^2 +a_0^2)^2 } (r-r_h), \quad A_r =0. \label{gf4r2}
\ee

The reader will note that the metric has an $\mathrm{AdS}_2$ factor, in the $t,r$ plane and an $\mathrm{SL}(2,\mathbb{R})$ symmetry with the generators given in 
eq.(\ref{sl2r}) with $L_2$ given by eq.(\ref{defL24}) and 
\begin{equation}
\label{4al}
\alpha = \frac{2  r_h a_0 \Xi_0}{(r_h^2 +a_0^2)^2 }.
\end{equation} 
The gauge field eq.(\ref{gf4r}) preserves the   $\mathrm{SL}(2,\mathbb{R})$ symmetry.

As discussed in \S\ref{ss-jtrot}, to calculate the leading dependence of the free energy at small temperatures, we need to keep the change,  due to non-extremality,
 in only some of the functions which appear in the metric, eq.(\ref{mt4}).  For example, the fractional change in $\rho^2$ is small in the whole of spacetime outside the horizon, similarly the change in  $\Delta_\theta$ and $\Xi$. Thus  the change in these function can be neglected and in the near-horizon region, we can take these function to have the form they do in the attractor geometry. 
 
 With these observations in mind, we  now take the near-horizon metric to have the form, 
\begin{equation}
\label{metan4}
\md s^2 = \frac{\rho_0^2}{(r_h^2 + a_0^2)} \left( \frac{\Phi_0}{\Phi} g_{\alpha \beta} \md x^\alpha \md x^\beta + \frac{\Phi^2 \Xi_0}{\Delta_{\theta 0} } \md \theta^2  \right) + \frac{\Delta_{\theta 0}(r_h^2 + a_0^2)}{\rho_0^2 \Xi_0} \Phi^2  \sin^2 \theta ( \md \varphi - \mathrm{i} A_\alpha \, \md x^\alpha )^2,
\end{equation}
and the gauge field to be of the form,
\begin{align}
\label{gfan4}
\hat{A}_\mu \, \md x^\mu = - \mathrm{i} \left( \mathcal{A}_\alpha + b(\theta) A_\alpha \right) \md x^\alpha + b(\theta) \, \md \varphi, 
\end{align}
where,
\begin{equation}
b(\theta) = \frac{1}{\Xi_0 \rho_0^2} \left(q_{e0} a_0 r_h \sin^2 \theta + q_{m0} (r_h^2 + a_0^2) \cos \theta  \right). \label{defb}
\end{equation}
Here, the fields $\Phi, A_\alpha, \mathcal{A}_\alpha$ are assumed to be dependent only on the 2-dimensional coordinates $x^\alpha$.
Note that volume of the two-sphere spanned by $\theta,\varphi$ is given by 
\be
\label{vols2}
V_2=4\pi \Phi^2.
\ee
 $\Phi_0$ appearing in eq.(\ref{metan4}) is the attractor value of $\Phi$. 
 
 Inserting in the action (\ref{dac4}) and carrying out the integral over $\theta,\varphi$ gives the two-dimensional action which agrees with the JT model,
  as is discussed in detail in  appendix \ref{app-rot4d}.   
 
 Here let us summarise the key point in the analysis in appendix \ref{app-rot4d}.   The $2$-dimensional action  which is obtained, eq.(\ref{euac2}), has the form,
 \be
 \label{tertwrot}
 I=-{1\over 4 G_{4}} \int \md^2 x \, \sqrt{g}\, \Phi^2 R  + \cdots .
 \ee
 Expanding $\Phi$, eq.(\ref{defphi}) then gives from the term linear in $\phi$ to be of the form eq.(\ref{JTact}) with ${\widetilde G}$ given in eq.(\ref{knag}). 
 From the expansion for $\Phi$ defined in eq.(\ref{phie}), we also learn that ${\cal J}$ is given by eq.(\ref{knaj}). 
 It then follows from the analysis in the JT model that the free energy, eq.(\ref{fJT}) agrees with the result obtained above, eq.(\ref{kfree}).

\subsection{The Near-Extremal Kerr Black Hole}
The asymptotically flat  uncharged Kerr black hole  is  an important special case.
This is obtained by taking   $L \rightarrow \infty$ and setting $q_e=q_m=0$ in all the  formulae in the previous section.

The parameter  $\mathcal{J}$, eq.(\ref{knaj}), takes the value 
\be
\label{jrh}
\mathcal{J} = \frac{1}{r_h}.
\ee
and, from eq.(\ref{knag}),  
\be
\label{valasf}
{\widetilde G}= \frac{G_{4}}{16\pi r_h^2}.
\ee
In this case, $a_0 = r_h$, eq.(\ref{a04}), and $r_h$ is related to the angular momentum of the black hole $J$ by 
\be
\label{valrh}
r_h= \sqrt{G_{4} J}.
\ee
Note that $r_h$ is of order the size of the event horizon.

The attractor values for $\Phi$, eq.(\ref{phiat4}) becomes, 
\be
\label{l2}
\Phi_0= \sqrt{2} r_h ,
\ee
and $\phi$ varies  as given in eq.(\ref{linvardil}) with ${\cal J}$ taking the value, eq.(\ref{jrh}).

The dimensional reduction discussion goes through as described above, resulting in a two-dimensional action of the form 
\be 
\label{nkf4jt}
I_{\mathrm{JT}} = - {1\over {16 \pi \widetilde G}} \left(  \int \md^2 x\, \sqrt{g} \, \phi \Big(R + \frac{2}{L_2^2}\Big) + 2 \int_{\del} \md x\, \sqrt{\gamma} \, \phi K   \right).
\ee
with ${\widetilde G}$ given in eq.(\ref{valasf}).

The condition for the temperature to be small, eq.(\ref{valT}),  using eq.(\ref{jrh}), becomes, 
\be
\label{condtt}
T\ll \mathcal{J}={1\over r_h}.
\ee
We see that the temperature has to be much smaller than the inverse ``light crossing time of the horizon''.

 From  eq.(\ref{jrh}) and eq.(\ref{valasf}), we see that the resulting free energy at small temperatures is
  \be
\label{fe4ff2}
F= - \frac{4\pi^2 r_h^3}{G_{4} } T^2 = -2 \pi S_0\sqrt{G_{4} J} T^2  = - 2 \pi S_0 \frac{T^2}{\mathcal{J}},
\ee 
 with $J$ being the angular momentum and $S_0$ being the zero-temperature entropy. This agrees with eq.(\ref{finalfree}) with $d=3$, see also \cite{Castro:2009jf}.
 
Note also that the chemical potential which is the thermodynamic conjugate of $J$ is given for an extremal Kerr black hole by, eq.(\ref{omhi4}),
\be
\label{valc}
\Omega_H={1\over 2 a_0}
 \ee
 while the angular momentum is given by, eq.(\ref{valrh}) 
 \be
 \label{angmom}
 J={a_0^2\over G_4}
 \ee
 
 This implies that the ``charge susceptibility'' \cite{Davison:2016ngz} at zero temperature is given by 
 \be
 \label{chargesus}
 {\md J\over \md \Omega_H}= - {4 a_0^3\over G_4} = -4 \sqrt{G_4 J^3}
 \ee
 
We had mentioned above in passing that an extra phase mode enters in the $2$-dimensional theory when one allows the charge to vary. It has an action 
 schematically of the form, $\int \md \tau  \mathcal{I} \dot{\theta}^2$. The susceptibility fixes the parameter $\mathcal{I}$, which acts like the moment of inertia of the phase mode. 
 
\subsection{Superradiant Instabilities}

Let us also briefly discuss superradiant instabilities which can occur in the asymptotic anti-de Sitter spacetime \cite{Cardoso:2004nk,Cardoso:2004hs,Cardoso:2006wa}; a discussion for the $5$-dimensional case is in appendix \ref{app-rot5d}; see also the end of section \ref{sec5drf}. 
A necessary condition for the superradiant instability to occur for black holes in asymptotically $\mathrm{AdS}_4$ spacetime is 
\cite{Hawking:1999dp,Cardoso:2006wa},
\begin{equation}
\label{sr4}
\Omega^\infty_H > \frac{1}{L}, 
\end{equation}
or equivalently, from eq.(\ref{omhi4}),
\begin{equation}
\label{srb4}
r_+^2 < a L.
\end{equation}
Note that the angular velocity $\Omega_H^\infty$ is  the thermodynamical variable  conjugate to the angular momentum $J$. 
From eq.(\ref{a04})  it is easy to see that  the inequality eq.(\ref{srb4}), is always met for uncharged  rotating extremal black holes ($q_e = q_m = 0$).
However,  it is possible to evade this bound for extremal black holes with sufficient amount of charge (see e.g., \cite{Caldarelli:1999xj} for related discussion).  

An extremal black hole is specified by three parameters, $q_e,q_m,a$ with the parameter $a$ satisfying the  condition,
\be
\label{condao}
a^2 < L^2.
\ee
The radius of the horizon,  $r_h$,  is obtained from eq.(\ref{a04}) and  given by 
\be
\label{valrha}
r_h^2={ -(L^2+a_0^2)+\sqrt{(L^2+a_0^2)^2+12L^2(a_0^2+q_0^2)}\over 6}.
\ee

To evade the superradiant instability for the extremal case, one needs,  
\begin{equation}
\label{din}
r_h^2 > a_0 L.
\end{equation} 

In the  supersymmetric case \cite{Kostelecky:1995ei}, (see also \cite{Caldarelli:1998hg}), $q_0$ is determined in terms of $a_0$ by  
\begin{equation}
\label{susyqo}
q_0^2 = a_0 L \left( 1+ \frac{a_0}{L} \right)^2.
\end{equation}
One finds  from eq.(\ref{valrha}) that in this case $r_h^2=a_0L$ and one is at the edge of meeting the condition, eq.(\ref{din}).
By increasing $q_0$ and making it bigger than the SUSY value, eq.(\ref{susyqo}), for fixed $a_0$, it is easy to see from eq.(\ref{valrha}) that $r_h$ increases and 
thus eq.(\ref{din}) is now met. 

In the SUSY case, if one makes the black hole slightly non-extremal, keeping the charges fixed at the SUSY values, one finds that   the inequality eq.(\ref{srb4}) is now violated and thus the superradiant instability goes away.


\section{Conclusions}
\label{sec-cocl}
In this paper, we have argued that the low-temperature free energy for near-extremal black holes is correctly obtained from the JT model in general.
In particular,  our  arguments  apply for rotating black holes as discussed in \S\ref{ss-jtrot}.
These conclusion holds for all cases where the extremal geometry has an $\mathrm{AdS}_2$ factor and the related $\mathrm{SL}(2,\mathbb{R})$  symmetry. 
 We have also verified by explicit calculations that these arguments are correct for near-extremal Kerr black hole in $5$  dimensions and for a general Kerr-Newman black hole in $4$ dimensions. 
 
 The dilaton field in the JT model corresponds to the volume of the transverse sphere spanned by the angular direction. 
 It follows from our analysis that the temperature dependence of the  free energy for all near-extremal black holes, including rotating ones,  is given in a compact form by eq.(\ref{finalfree}) and determined by the ground state entropy and the scale ${\cal J}$ which characterises  how the dilaton approaches its attractor value at the horizon.

The expectation that the JT model should be a good approximation, at sufficiently low temperatures and low frequencies  for rotating black holes, which have an $\mathrm{AdS}_2$ factor in their near-horizon geometry, already arises from previous work \cite{Astefanesei:2006dd,Moitra:2018jqs}.  In \cite{Astefanesei:2006dd}, it was argued that extremal rotating black holes would also exhibit the attractor behaviour.   One way to understand this is to construct the two-dimensional theory obtained after reducing over the angular direction. This theory contains the two-dimensional metric, the dilaton, which is the volume of the transverse sphere and appears as the coefficient of the Einstein-Hilbert term in the two-dimensional theory,  several gauge fields, scalars and also massive symmetric rank two tensor fields. In the rotating case, some of the gauge fields which are excited arise from components of the higher dimensional metric and similarly 
some of the scalars and the massive symmetric rank two fields arise from partial waves on the transverse sphere -- these are  excited due to the rotation. 
However, despite these complications, the analysis in the resulting two-dimensional theory is similar  in the rotating and non-rotating cases, and all fields are drawn to their attractor values which can be obtained by extremising an entropy function. 

The analysis of \cite{Astefanesei:2006dd} was extended in \cite{Moitra:2018jqs} to include the slightly non-extremal case. It was argued that while allowing the dilaton to vary from its attractor value is important for  controlling the back reaction, the departure for many other fields could be neglected to leading order, at low frequencies and low temperatures. 
This gives rise to a theory consisting of the JT model coupled to extra phase modes which arise from the gauge fields present in the two-dimensional theory. It follows from the arguments  in \cite{Moitra:2018jqs} that this theory should be a good approximation for the low-frequency and low-temperature response of a near-extremal black hole. While rotating black holes were not explicitly considered in \cite{Moitra:2018jqs}, the arguments there were based on the two-dimensional theory obtained after dimensional reduction over the transverse sphere and should apply for the rotating case as well, since as mentioned above, the resulting two-dimensional theories are similar. 

In this paper, we have focussed on the low-temperature behaviour for near-extremal rotating black holes and confirmed both through a general argument and also explicit checks that this expectation is correct and that the JT model does  indeed correctly account for the low-temperature free energy of such rotating near-extremal black holes.

 These developments hold out the exciting possibility that the JT model could, in fact, be useful in analysing the dynamics of fast-spinning near-extremal black holes found in nature!  
In the paper, we saw that the thermodynamics of the black hole is correctly reproduced if the dilaton satisfies Dirichlet boundary conditions at the  boundary of the $\mathrm{AdS}_2$ region. From the point of view of the four-dimensional metric, this means taking the volume of the transverse sphere to be constant along the boundary. Therefore, these results suggest that the dynamics of near-extremal rotating black holes, at sufficiently low frequencies but not necessarily small amplitudes,  could be well approximated by an effective theory in which the near-horizon $\mathrm{AdS}_2$ region is  replaced by a boundary located along the locus where  the transverse sphere takes a constant volume. This boundary  would interact with the stress tensor averaged over the transverse sphere and the dynamics of the black hole would  then  arise through these from  interactions of the boundary.\footnote{As extra phase mode would also enter in the boundary action, since the in falling matter could change the angular momentum of the black hole. Its moment of inertia is determined by the charge susceptibility 
given in eq.(\ref{chargesus}).}
 
    To get a feel for some of the numbers involved, consider a solar mass extremal Kerr black hole. It has a horizon of approximately $3$ km, with a light crossing time of $10^{-5}$ sec. Thus, one would expect that the JT model could  be a good approximation for dynamics at frequencies which are smaller than  about $10^5$ Hz.     Some examples of  fast-spinning black holes are well known with  masses  in the range of a few solar masses. For example, GRS 1915+105 involves a binary system with a black  hole of mass $\sim 10 M_{\odot} $ and a spin parameter which has been measured to be around $0.98$ \cite{McClintock:2006xd, Blum:2009ez, Miller:2013rca} -- making it very nearly extremal. 
        
 We hope to develop this connection with astrophysical fast-spinning black holes further in subsequent work. 
 
 The analysis in this paper and also in \cite{Nayak:2018qej} and \cite{Moitra:2018jqs}, suggest that one should be able to formulate a version of the fluid-gravity correspondence for near-extremal black holes including rotating ones.\footnote{We are grateful to ``The Professor'' -- Shiraz Minwalla -- for his comments in this regard.} Consider a near-extremal black hole in asymptotically $\mathrm{AdS}$ spacetime with  temperature $T\ll \mu$, where $\mu$ is the chemical potential corresponding to the charge carried by the black hole.
   For situations where 
   $L^{-1} \ll T$,  with $L$ being the length or time scale over which  the fluid in the boundary theory varies, it is well known  \cite{Bhattacharyya:2008jc} that Einstein's equations lead to Navier-Stokes equations in the boundary theory. One might expect that  when the variation is more rapid and satisfies the condition $T\ll L^{-1} \ll \mu$,
 also  there is  an effective theory that describes the dynamics, and it is obtained by coupling the JT model at the boundary of $\mathrm{AdS}_2$ in a suitable way with the outside.  We hope to develop this theme in a subsequent paper as well.

The universal manner in which the JT model describes near-extremal black holes also allows for a fairly precise near-$\mathrm{AdS}_2$/near-$\mathrm{CFT}_1$ correspondence to be set up \cite{Moitra:2018jqs}. It will be very interesting to test this correspondence in  examples where the extremal black holes arise in  in  string theory, with a near-$\mathrm{CFT}_1$ which can be understood in a microscopically precise way. The fact that rotating black holes are included in these examples, enlarges the set of possibilities even further. It might well be that some  instances where this correspondence can be tested and understood most fruitfully involve rotating black holes.

\acknowledgments
We thank   Indranil Halder and  Shiraz Minwalla,  for insightful discussions and comments. We thank the DAE, Government of India, for support.  
SPT also acknowledges support from the J. C. Bose fellowship of the DST, Government of India. 
UM thanks the Israel Institute for Advanced Studies, The Hebrew University of Jerusalem, Israel ({\it The 36th Advanced School in Physics on Recent Progress in Quantum Field/String Theory}) and the Abdus Salam International Centre for Theoretical Physics, Italy ({\it Spring School on Superstring Theory and Related Topics}) and the other organisers of these programmes for their warm hospitality during different stages of this work. 
We also acknowledge support from the Infosys Endowment for Research on the Quantum Structure of Spacetime. 
Most of all, we thank the people of India for generously supporting research in String Theory.

\appendix


\section{On How the Jackiw-Teitelboim Model Correctly Reproduces the Leading Thermodynamic Behaviour}
\label{app-jttd}

In this appendix, we elaborate on the comments made in section \ref{ss-jtm} and show exactly how the evaluation of the Euclidean on-shell action in the region far from the horizon  (the bulk integral extending from the $\mathrm{AdS}_2$ boundary to the asymptotic infinity) gives rise to the Gibbons-Hawking boundary term and the counter-term in the Jackiw-Teitelboim model -- which are precisely the terms contributing to the JT on-shell action. To begin we consider the RN case studied in 
\ref{ssec-mdrn}. The integral of interest in given in eq.(\ref{vals2}),

\begin{align}
I_{[\partial \mathrm{AdS}_2\rightarrow \infty]} &= - \frac{1}{4G_{4} } \int_{\del \mathrm{AdS}_2}^\infty \md^2 x \, \sqrt{g} \left(  \Phi^2 R + \frac{2\Phi_0}{\Phi} - 2\Phi_0 \Phi \Lambda + \frac{\Phi^3}{\Phi_0} F^{\alpha \beta} F_{\alpha \beta}  \right) \nonumber \\  
&\quad - \frac{1}{2G_{4} } \int_{\mathcal{\infty}} \md x \, \sqrt{\gamma}\, \left( \Phi^2 K - \frac{2\Phi^3}{\Phi_0} n_\alpha F^{\alpha \beta} A_\beta \right)  + I_{\mathrm{CT}}^\infty  \label{avals2} \\
&\quad - \frac{1}{G_{4} } \int_{\partial \mathrm{AdS}_2} \md x\, \sqrt{\gamma} {\Phi^3\over \Phi_0} n_{\alpha} F^{\alpha \beta} A_\beta . \nonumber 
\end{align}
We emphasise again that the normal one-form $n_\alpha$ appearing in the last line above (and the discussion below) points from the $\mathrm{AdS}_2$ boundary to the asymptotic infinity.

The change in $I_{[\partial \mathrm{AdS}_2\rightarrow\infty]}$ to first order in the metric, dilaton and gauge field perturbations, which we denote by,  $\delta g_{\alpha\beta}$,  $\delta \Phi$ and 
$\delta A_\alpha$, respectively, is given by, 
\begin{align}
\delta I_{[\partial \mathrm{AdS}_2\rightarrow \infty]} &= 
- \frac{1}{4G_{4} } \int_{\del \mathrm{AdS}_2} \md x \, \sqrt{\gamma} \, \Big(  \Phi^2 n_\beta \nabla_\alpha \delta g^{\alpha \beta}  -  \Phi^2 g_{\alpha \beta} n^\kappa \nabla_\kappa \delta g^{\alpha \beta}  \Big)\nonumber \\
&\quad + \frac{1}{4G_{4} } \int_{\del \mathrm{AdS}_2} \md x \, \sqrt{\gamma} \, \delta g^{\alpha \beta}  \Big( n_\beta \nabla_\alpha \Phi^2 - g_{\alpha \beta} n^\kappa \nabla_\kappa \Phi^2 \Big)  \label{viai}  \\
&\quad  - \frac{1}{G_{4} } \int_{\partial \mathrm{AdS}_2} \md x \, \delta \left(\sqrt{\gamma} {\Phi^3\over \Phi_0} n_{\alpha} F^{\alpha \beta} \right) A_\beta , \nonumber 
\end{align}
where we have used the equations of motion, and also the fact that all fields vanish sufficiently fast at infinity.
Note that all the three terms are surface terms at the boundary of $\mathrm{AdS}_2$, in agreement with the discussion in section \ref{ss-jtm}. Note also that the quantity that is being varied in the third line of eq.(\ref{viai}) above is proportional to the charge carried by the black hole (see eq.(\ref{fs})). Since we are  interested in perturbations which do not change this charge, the term in the last line of eq.(\ref{viai}) vanishes.  

In order to obtain the first two lines of eq.(\ref{viai}), we have used the Palatini identity,
\begin{equation}
\delta R_{\alpha \beta} = \nabla_\kappa \delta \Gamma^\kappa_{\alpha \beta} - \nabla_\alpha \delta \Gamma^\kappa_{\beta \kappa},
\end{equation} 
where the variation of the Christoffel connection is given by,
\begin{equation}
\delta \Gamma^\lambda_{\alpha \beta} = \frac{1}{2} g^{ \lambda \kappa} \left( \nabla_\alpha \delta g_{\beta \kappa} + \nabla_\beta \delta g_{\alpha \kappa} - \nabla_\kappa \delta g_{\alpha\beta }\right).
\end{equation}
The covariant differentiation and rearrangement of indices above are assumed to be done with the unperturbed metric. 

As discussed in section \ref{sec-rev-sph}, for obtaining the leading dependence of the action on temperature, it is enough to consider only terms up to linear order in $\phi$, 
eq.(\ref{defphi}) at the boundary of $\mathrm{AdS}_2$. Expanding eq.(\ref{viai}) up to linear order in $\phi$ we get, 
\begin{align}
\delta I_{[\del \mathrm{AdS}_2 \to \infty]} &= - \frac{1}{16\pi \widetilde{G} } \int_{\del \mathrm{AdS}_2} \md x \, \sqrt{\gamma} \, \Big(  \phi n_\beta \nabla_\alpha \delta g^{\alpha \beta}  -  \phi g_{\alpha \beta} n^\kappa \nabla_\kappa \delta g^{\alpha \beta}  \Big)  \label{varai} \\
&\quad + \frac{1}{16\pi \widetilde{G} } \int_{\del \mathrm{AdS}_2} \md x \, \sqrt{\gamma} \, \delta g^{\alpha \beta}  \Big( n_\beta \nabla_\alpha \phi - g_{\alpha \beta} n^\kappa \nabla_\kappa \phi \Big) \nonumber.
\end{align}
Here we have dropped some terms coming from the first line on the RHS of eq.(\ref{viai}) which are  independent of $\phi$ and turn out to vanish.

Let us now look at the Gibbons-Hawking boundary term on the $\mathrm{AdS}_2$ boundary, given by,
\begin{equation}
\label{agh}
I_{\mathrm{GH} } = -\frac{1}{8\pi \widetilde{G} }\int_{\del \mathrm{AdS}_2} \md x \, \sqrt{\gamma} \,\phi K.
\end{equation}
Its variation gives,
\begin{equation}
\label{aghv}
\delta I_{\mathrm{GH} } = -\frac{1}{8\pi \widetilde{G} } \int_{\del \mathrm{AdS}_2} \md x \, \delta(\sqrt{\gamma}) \,\phi K -\frac{1}{8\pi \widetilde{G} }\int_{\del \mathrm{AdS}_2} \md x \, \sqrt{\gamma} \,\phi \, \delta K.
\end{equation}
Note that, crucially, we have imposed a strict Dirichlet boundary condition on $\phi$ on the $ \mathrm{AdS}_2$ boundary. We now use the relation that,
\begin{equation}
\label{delK}
\delta K = \nabla_\alpha \delta n^\alpha - \frac{1}{2} n^\alpha g_{\kappa \lambda} \nabla_\alpha \delta g^{\kappa \lambda},
\end{equation}
where,
\begin{equation}
\label{dna}
\delta n^\alpha = - \frac{1}{2} n^\alpha n_\kappa n_\lambda \delta g^{\kappa \lambda} + \delta g^{\alpha \beta} n_\beta.
\end{equation}

Using these relations, we obtain, after some simplification,
\begin{align}
\delta I_{[\del \mathrm{AdS}_2 \to \infty]} - \delta I_{\mathrm{GH} } &= \frac{1}{16 \pi \widetilde{G} } \int_{\del \mathrm{AdS}_2} \md x \, \sqrt{\gamma} \Big( \nabla_\alpha ( \phi n_\beta \delta g^{\alpha \beta} ) + \delta g^{\alpha \beta} \phi (\nabla_\alpha n_\beta)  \Big) \nonumber \\
&\quad + \frac{1}{16 \pi \widetilde{G} } \int_{\del \mathrm{AdS}_2} \md x \, \sqrt{\gamma} \Big( \phi \nabla_\kappa (  n^\kappa \gamma_{\alpha \beta} \delta g^{\alpha \beta} )- \nabla_\kappa (\phi n^\kappa g_{\alpha \beta} \delta g^{\alpha \beta} )  \Big) \label{vdf} \\
&\quad + \frac{1}{8\pi \widetilde{G} } \int_{\del \mathrm{AdS}_2} \md x \, \delta(\sqrt{\gamma}) \,\phi K , \nonumber
\end{align}
where,
\begin{equation}
\label{defin}
\gamma_{\alpha \beta} = g_{\alpha \beta} - n_\alpha n_\beta.
\end{equation}

To proceed, we use a coordinate system, with $r,t$ coordinates, in which the metric of $\mathrm{AdS}_2$ is 
\begin{equation}
\label{agt}
g_{tt} = \frac{1}{g_{rr}} = \frac{(r-r_h)^2}{L_2^2},\quad  g_{tr} =0,
\end{equation}
and the dilaton, eq.(\ref{valphia}) becomes, 
\begin{equation}
\label{pr}
\phi = {r -  r_h \over r_h}.
\end{equation}
This gives on the constant $r$ hyper-surface on the $\mathrm{AdS}_2$ boundary, the unit normal one-form,
\begin{equation}
\label{nr}
n_r = \frac{L_2}{r-r_h}, \quad n_t  = 0.
\end{equation}

In this coordinate system  the metric components, eq.(\ref{agt}), change as follows, when the mass $M$, eq. (\ref{valf}) is changed by $\delta M$ to go to the non-extremal case:
\begin{equation}
\label{mdf}
\delta g^{tt} = \frac{2 L_2^4 G_{4} \delta M }{r_h (r-r_h)^4 }, \quad \delta g^{rr} = - \frac{2 G_{4} \delta M }{r_h}, \quad \delta g^{tr} = 0.
\end{equation}
Using this explicit form, we find that the first two lines of (\ref{vdf}) vanish and we are left with,
\begin{equation}
\label{vab1}
\delta I_{[\del \mathrm{AdS}_2 \to \infty]} - \delta I_{\mathrm{GH} } =  \frac{1}{8\pi \widetilde{G} } \int_{\del \mathrm{AdS}_2} \md x \, \delta(\sqrt{\gamma}) \,\phi K.
\end{equation}
Since $K = 1/L_2$, we have,
\begin{equation}
\label{vab2}
\delta I_{[\del \mathrm{AdS}_2 \to \infty]} = \delta I_{\mathrm{GH} } +  \frac{1}{8\pi \widetilde{G} L_2} \int_{\del \mathrm{AdS}_2} \md x \, \delta(\sqrt{\gamma}) \,\phi .
\end{equation}

We thus see that the change to leading order in $I_{[\del \mathrm{AdS}_2 \to \infty]}$ agrees with the change in  the Gibbons-Hawking boundary term and the associated counter-term on the $\mathrm{AdS}_2$ boundary.

Before we end this appendix, let us make two comments. 
First, with reference to the discussion in \ref{ss-jtrot}, instead  of starting with the action, eq.(\ref{avals2}), consider the linearised action of the JT model, eq.(\ref{bulkJT}) which is now evaluated in the  region extending from the horizon to the boundary of $\mathrm{AdS}_2$. And let us take an alteration of the metric which can be regarded as a small perturbation in this region.  Then steps which are very analogous to those above lead to a change given in eq.(\ref{changeact}).  

Second, consider the general two-dimensional theory of \S\ref{ss-jtgt}. All the above arguments go through as before. The new ingredient in this case is the presence of the extra scalars $\Psi_p$. When we vary the far-horizon action, some surface terms are generated on the $\mathrm{AdS}_2$ boundary. The relevant term in the action (\ref{actag}) is,
\begin{equation}
\label{s1}
- \frac{v_{d-1}}{16\pi G_{d+1} } \int \md^2 x \, \sqrt{g} \,  h^{pq}(\Phi, \Psi_i) g^{\alpha \beta} 
  \nabla_\alpha \Psi_p \nabla_\beta \Psi_q .
\end{equation}
A variation of this action yields,
\begin{equation}
\label{vs1}
 -\frac{v_{d-1}}{8\pi G_{d+1} } \int_{\del \mathrm{AdS}_2} \md x \, \sqrt{\gamma} \, h^{pq} n^\alpha 
  \del_\alpha \Psi_p  \delta \Psi_q .
\end{equation}

Unlike in the case of $\Phi$, no Dirichlet condition on $\Psi$ is imposed at the $\mathrm{\mathrm{AdS}}_2$ boundary. Therefore, the above expression does not vanish, in general. However, it is easy to see that this does not contribute to the leading thermodynamic behaviour we are interested in. Recall from \S\ref{ss-jtgt} that $\Psi_i$ also shows the behaviour (\ref{scexp}).  In the asymptotic $\mathrm{AdS}_2$ region the effects of the mass deformation die out and this implies that  
$\delta \Psi_i$  will also go as  $\mathcal{J}_i^{-1}$.  The net contribution of this boundary term will therefore be  suppressed  at least as $\mathcal{J}_i^{-2}$. Since $\mathcal{J}$ and $\mathcal{J}_i$ are assumed to be of similar order, this contribution is suppressed in comparison to the leading thermodynamic behaviour which goes as $\mathcal{J}^{-1}$.

Let us comment on the behaviour of $\delta \Psi_i$ in the asymptotic $\mathrm{AdS}_2$ region in some more detail.  Consider the mass deformed two-dimensional metric which can be written as, 
\be
\label{massdef}
\md s^2= (r^2-\delta_0^2) \, \md t^2+{\md r^2\over (r^2-\delta_0^2)}.
\ee
Here we have relabelled to the coordinate $(r-r_h)$ used in section \ref{sec-rev-sph} as $r$ and $\delta_0 \sim T$ .
A  perturbation $\delta \Psi_i$ of mass $m$ which is time-independent satisfies the equation 
\be
\label{eomperaa}
\partial_r[(r^2-\delta_0^2) \partial_r \delta \Psi_i] -m^2 \delta \Psi_i=0.
\ee
In the asymptotic $\mathrm{AdS}_2$ region, $r^2\gg \delta_0^2 \sim T^2$,  eq.(\ref{eomperaa})  becomes the same as in the undeformed case. Regularity at the horizon chooses one of the two solutions of eq.(\ref{eomperaa}); the remaining overall coefficient which specifies  this solution is fixed by matching with the behaviour in the asymptotic $\mathrm{AdS}_2$ region. This allows the solution to meet the boundary conditions at infinity. 
For the $m^2=2$ case, for instance,  the behaviour in the asymptotic $\mathrm{AdS}_2$ region goes like $\sim {r/{\cal J}_i}$, and in fact, 
\be
\label{colperaa}
\delta \Psi_i= {r\over L_2^2  {\cal J}_i}
\ee
solves eq.(\ref{eomperaa}) identically. 
As the value of the dilaton in the asymptotic region changes due to the mass deformation, the boundary of the $\mathrm{AdS}_2$ region also changes, however $\delta \Psi_i$ will continue to be of order ${1/ {\cal J}_i}$.

\section{Electrically Charged Near-Extremal Black Holes in $(d+1)$ Dimensions}
\label{app-gd}

The electrically charged RN case in general $d+1$ dimensions has been studied earlier, \cite{Sachdev:2019bjn,Moitra:2018unp}. Here we review how the free energy for this case agrees with the JT model.

We consider Euclidean Einstein-Maxwell action in asymptotically $\mathrm{AdS}_{d+1}$ spacetime, appropriate for the canonical ensemble,
  \be 
  \label{ehdd}
I=  -\frac{1}{16\pi G_{d+1}} \int \md^{d+1} x\, \sqrt{g} \, \left( R - 2 \Lambda_{d+1} + F_{\mu \nu}  F^{\mu \nu} \right)   - \frac{1}{8\pi G_{d+1}} \int  \md^{d} x\, \sqrt{\gamma}\, \left( K - 2 n_\mu F^{\mu \nu} A_\nu \right) .
  \ee

The electrically charged Reissner-Nordstr\"{o}m solution is given by,
\begin{equation}
\label{mdp1}
\md s^2 =  f(r) \, \md t^2 +\frac{\md r^2}{f(r)} + r^2 \, \md \Omega_{d-1}^2 ,
\end{equation}
where,
\begin{align}
f(r) &= \left( 1 - \frac{2 m}{r^{d-2}} + \frac{q^2}{r^{2d-4}} + \frac{r^2}{L^2} \right),  \label{dfr} \\
 L  &= \sqrt{-\frac{d(d-1)}{2 \Lambda_{d+1} } }, \label{ladd} \\
 F_{rt} &= \sqrt{  \frac{(d-2)(d-1)} {2} } \frac{q}{r^{d-1}}. \label{frtd}
\end{align}

The physical mass is given by,
\begin{equation}
\label{pmd}
M  = M_{\mathrm{Casimir}} + \frac{(d-1) v_{d-1} m}{8\pi G_{d+1}}.
\end{equation}
Here we have defined
\begin{equation}
\label{dfvd}
v_{d} \equiv \mathrm{Vol}(S^d) = \frac{2\pi^{\frac{d+1}{2}}}{\Gamma(\frac{d+1}{2}) }.
\end{equation}
We have, in general, a non-zero $M_{\mathrm{Casimir}}$ when $d$ is even.

In the  extremal case $f(r)$, eq.(\ref{dfr}) has a double zero at the horizon, $r=r_h$. Increasing the mass by $\delta M$ keeping the charge parameter $q$ fixed, 
 we find the near-horizon  metric to be,
\begin{equation}
\md s^2 =  \frac{(r-r_h)^2 - \delta_0^2}{L_2^2} \md t^2 + \frac{L_2^2}{(r-r_h)^2 - \delta_0^2} \md r^2 + r^2 \, \md \Omega_{d-1}^2, \label{nhd}
\end{equation}
where,
\begin{align}
\frac{1}{L_2^2} &= \frac{d(d-1)}{L^2} + \frac{(d-2)^2}{r_h^2}, \label{l2d}  \\
\delta_0^2 &= \frac{16 \pi G_{d+1} L_2^2 \, \delta M}{ (d-1) v_{d-1} r_h^{d-2} }. \label{dod} 
\end{align}

It is easy to see that the temperature is given by 
\be
\label{valtaa}
T= \frac{\delta_0}{2\pi L_2^2},
\ee
and the entropy  above extremality is given by 
\be
\label{entaa}
\Delta S=\frac{\pi (d-1) v_{d-1} L_2^2 r_h^{d-2}}{2 G_{d+1} } T.
\ee
This leads to a free energy (using the usual thermodynamic relation), 
\begin{equation}
\label{tfsd}
\Delta F = - \frac{\pi (d-1) v_{d-1} L_2^2 r_h^{d-2}}{4 G_{d+1} } T^2 .
\end{equation}

Next let us compare this with the result in the JT model obtained in a manner analogous to what was described  in section \ref{sec-rev-sph} for the four-dimensional case.
Dimensionally reducing  the action (\ref{ehdd}) with the ansatz,
\begin{equation}
\label{and}
\md s^2 =  \left(\frac{\Phi_0}{\Phi} \right)^{d-2} g_{\alpha \beta} \, \md x^\alpha \, \md x^\beta + \Phi^2 \, \md  \Omega_{d-1}^2.
\end{equation} 
gives the two-dimensional action,
\begin{align}
I &= - \frac{v_{d-1} }{16\pi G_{d+1} }\int \md^2 x \, \sqrt{g} \left( \Phi^{d-1} R + \frac{(d-1)(d-2) \Phi_0^{d-2} }{\Phi} +  \frac{d(d-1)}{L^2} \Phi_0^{d-2} \Phi + \frac{\Phi^{2d-3}}{\Phi_0^{d-2}} F_{\alpha \beta} F^{\alpha \beta} \right) \nonumber \\
&\quad - \frac{v_{d-1} }{8\pi G_{d+1} } \int \md x \, \sqrt{\gamma} \left( \Phi^{d-1} K - \frac{2\Phi^{2d-3}}{\Phi_0^{d-2}} n_\alpha F^{\alpha \beta} A_\beta \right). \label{2dda}
\end{align}
Here 

\begin{equation}
\label{pdd}
\Phi = r,
\end{equation}
and $\Phi_0 = r_h$ is the attractor value for $\Phi$. Expanding about the attractor value we see that,
\begin{equation}
\label{pdv}
\Phi = \Phi_0 \left(1 + \frac{r - r_h}{r_h} \right)= \Phi_0 \left(1 + \frac{1}{\mathcal{J} z } \right).
\end{equation}
where ${\cal J}$, which determines the scale of conformal symmetry breaking in the near-horizon region eq.(\ref{linvardil}), is 
\begin{equation}
\label{mjd}
\mathcal{J} = \frac{r_h}{L_2^2}.
\end{equation}

We get from the equations of motion,
\begin{equation}
\label{daf}
F_{\alpha \beta} = \frac{\Phi_0^{d-2}}{\Phi^{2d-3}} Q \sqrt{g} \epsilon_{\alpha \beta},
\end{equation}
where,
\begin{equation}
Q^2 =  \frac{1}{2}(d-1)(d-2)  q_0^2, \label{qd}
\end{equation}
with $q_0$ referring to the charge parameter at extremality. 

Following the procedure outlined in section \ref{ss-jtm} and expanding about the attractor value of $\Phi$, we readily obtain the bulk action,
\begin{align}
I=- \frac{v_{d-1} (d-1) r_h^{d-1} }{16 \pi G_{d+1} }  \int \md^2 x \, \sqrt{g} \, \phi \left(R+ \frac{2}{L_2^2} \right) . \label{JTd}
\end{align}

This agrees with the JT model bulk action in eq.(\ref{JTact}) and allows us to identify,
\begin{equation}
\label{geffd}
\widetilde{G} = \frac{G_{d+1} }{(d-1) v_{d-1} r_h^{d-1}}  .
\end{equation}

From eq.(\ref{fJT}), eq.(\ref{mjd}) and eq.(\ref{geffd})  we see that the free energy of this JT model also agrees with eq.(\ref{tfsd}) above.

\section{Rotating Black Holes in Five Dimensions in Asymptotically Anti-de Sitter Spacetime}
\label{app-rot5d}

In this appendix, we give further details on the five-dimensional rotating black hole, discussed in section \ref{sec5drf}. The five-dimensional action is given by eq.(\ref{n5deinac}) in which the five-dimensional cosmological constant is given by,
\begin{equation}
\label{5dccdef}
\Lambda_{5} =  - \frac{6}{L^2}.
\end{equation}

The five-dimensional rotating black hole solution \cite{Castro:2018ffi} is given by eq.(\ref{5met}), where,
\begin{align}
\Xi & =1-\frac{a^2}{L^2}, \label{kads5-Xi} \\
\Delta &=1+\frac{r^2}{{L}^2}-\frac{2 m r^2}{(a^2+r^2)^2}, \label{kads5-Del} \\
\exp(-U_1) &= \frac{a^2+r^2}{4 \Xi}, \label{kads5-U1} \\
\exp(-U_2) &= \frac{ma^2 }{2 \Xi ^2 (a^2+r^2)}+\frac{a^2+r^2}{4 \Xi }, \label{kads5-U2} \\
A &=\cos \theta \, \md \varphi  - \mathrm{i} \frac{a}{2\Xi} \left(\frac{a^2+r^2}{L^2}-\frac{2 m}{a^2+r^2}\right) \exp(U_2) \, \md t. \label{kads5-At-met}
\end{align}

The physical mass $M$ and the angular momentum are given by, respectively,
\begin{align}
M &= \frac{3\pi L^2 }{32 G_{5} } + \frac{3\pi \left(  1 + \frac{a^2}{3L^2} \right)}{4 G_{5} \Xi^3} m,  \label{adm5}\\
J &= \frac{\pi m a}{G_{5} \Xi^3}. \label{j5}
\end{align}
Here, the first term in (\ref{adm5}) corresponds to the Casimir energy for this spacetime. The entropy and the temperature are given respectively by,
\begin{align}
S &= \frac{\pi^2 (r_+^2 + a^2)^2 }{2 G_{5} r_+ \Xi^2 }, \label{ent5}\\
T &= \frac{r_+^2 -a^2 + \frac{2r_+^4}{L^2}}{2\pi r_+ (r_+^2 + a^2)}, \label{te5}
\end{align}
where $r_+$ denotes the location of the outer horizon. Note that the above formulae make it obvious that the absolute value of the parameter $a$ cannot be arbitrarily large. It can, at best, approach $L$ from below.  

The angular velocity which is thermodynamic conjugate of the angular momentum  \cite{Gibbons:2004ai} is given by,
\begin{equation}
\label{omhi5}
\Omega_H^\infty = a  \frac{\left( 1 + \frac{r_+^2}{L^2} \right)}{r_+^2 + a^2}.
\end{equation}

At extremality, we have the relations,
\begin{align}
m_{0} &= 2 r_h^2 \left( 1+ \frac{r_h^2}{L^2} \right)^3, \label{kads5-mext}\\
a_{0} &= r_h \left( 1+ \frac{2r_h^2}{L^2} \right)^{1/2}. \label{kads5-aext}
\end{align}
The fixed angular momentum condition gives us,
\begin{align}
m &= m_{0} + \delta m,  \label{kads5-mdefo}\\
a &= a_{0}  - \frac{\left( 1- \frac{2r_h^2}{L^2} \right) \left( 1+ \frac{2r_h^2}{L^2} \right)^{1/2}}{\left( 1+ \frac{r_h^2}{L^2} \right)^2 \left( 1+ \frac{5r_h^2}{L^2}+\frac{10r_h^4}{L^4} \right)} \frac{\delta m}{2r_h}. \label{kads5-adefo}
\end{align}
The horizon is shifted to,
\begin{equation}
r_+ = r_h + \delta_0 + \mathcal{O} ( \delta_0^2) ,
\end{equation}
where
\begin{equation}
\label{kads5-del0}
\delta_0^2 = \frac{3 \, \delta m}{2} \frac{\left( 1+ \frac{2r_h^2}{3L^2} \right) }{\left( 1+ \frac{4r_h^2}{L^2} \right) \left( 1+ \frac{5r_h^2}{L^2}+\frac{10r_h^4}{L^4} \right)}.
\end{equation}

The extra mass above extremality and the temperature are related as,
\begin{equation}
\label{kads5-M-T}
\delta M = \frac{2 \pi^3 r_h^2 L_2^2}{G_{5} } \frac{\left(1 - \frac{r_h^2}{L^2} \right)}{\left(1 + \frac{r_h^2}{L^2} \right)\left(1 - \frac{2r_h^2}{L^2} \right)^2}  T^2.
\end{equation}
The $L_2$ used here is defined in eq.(\ref{valads25}). At small temperature, the excess entropy is found to be,
\begin{equation}
\label{kads5-stj}
\Delta S = \frac{12\pi^3 r_h^3}{  G_{5} \left(1 - \frac{2r_h^2}{L^2} \right)^2} \frac{T}{\mathcal{J}}, 
\end{equation}
From which we easily obtain the form of free energy as in eq.(\ref{nhf}), where $\widetilde{G}, \mathcal{J}$ are defined in eqs.(\ref{valtg5},\ref{valj5}).

A criterion for superradiance instability to occur \cite{Hawking:1999dp} is given by,
\begin{equation}
\label{sr5}
\Omega^\infty_H > \frac{1}{L}. 
\end{equation}
Since $a$ can approach $L$ only from below, it it easy to see that this criterion is equivalent to demanding that,
\begin{equation}
\label{srb5}
r_+^2 < a L.
\end{equation}
In the purely rotating five-dimensional near-extremal case, one finds  that this criterion is  always met and hence an instability could exist \cite{Hawking:1999dp}.

\subsection{Dimensional Reduction to Two Dimensions}\label{subsec-5d-2d}

We now perform the dimensional reduction of the action (\ref{n5deinac}) -- first with the metric ansatz (\ref{met4}) to obtain the four-dimensional action (\ref{actfd}), then with the metric and gauge field ansatz (\ref{fourdmet}, \ref{gfa5}). The resulting two-dimensional action becomes, after we have added a gauge field boundary term for the canonical ensemble,

\begin{align}
I &= -\frac{\pi}{8G_{5} } \int \md^2 x\, \sqrt{g} \left( \Phi^3 R + \frac{16}{\Phi^{3/2}}   - \frac{3 \Phi^3}{2\Sigma^2} (\nabla \Sigma)^2 - \frac{2\Phi^{3/2} }{\Sigma} \Lambda_{5} - \frac{32\Sigma^3}{\Phi^{9/2}} + \frac{\Phi^{9/2} \Sigma^3}{4} F_{\alpha \beta} F^{\alpha \beta} \right) \nonumber \\
&\quad -\frac{\pi}{4G_{5} } \int \md x\, \sqrt{\gamma} \left( \Phi^3 K  - \frac{\Phi^{9/2} \Sigma^3}{2} n_\alpha F^{\alpha \beta} A_\beta \right) . \label{kads5-2dac}
\end{align}

The equation of motion for the gauge field gives,
\begin{equation}
\label{kads-f}
F_{\alpha \beta} =  \frac{Q}{\Phi^{9/2} \Sigma^3} \, \sqrt{g} \epsilon_{\alpha \beta},
\end{equation}
where the charge $Q$ is given by,
\begin{equation}
\label{kads-q}
Q^2 = 32 \Sigma_0^3 \Phi_0^3 \left( 1 - \frac{\Phi_0^3}{8 \Sigma_0} \Lambda_{5} - \frac{2\Sigma_0^3}{\Phi_0^3}  \right) = \pqty{ \frac{4G_5}{\pi} J_0 }^2.
\end{equation}
Here, $J_0$ is the extremal value of the angular momentum (\ref{j5}) and  $\Phi_0, \Sigma_0$ are the attractor values of the corresponding scalars. Comparing with the attractor solution (\ref{5attg}), we readily find that
\begin{align}
\Sigma_0 &= \frac{r_h}{ \left(1 - \frac{2r_h^2}{L^2} \right) }, \label{kads52S0}
\end{align}
with the value of $\Phi_0$ given by eq.(\ref{attracphi}).

Now, following the general procedure discussed in section \ref{ss-jtm}, and linearising about the attractor solutions, $\Phi = \Phi_0 (1+ \phi)$ and $\Sigma = \Sigma_0 (1+ \sigma)$, we have the bulk action in the JT form, 
 
\begin{align}
I = - \frac{3\pi r_h^3}{2 G_{5} \left(1 - \frac{2r_h^2}{L^2} \right)^2} \int \md^2 x\, \sqrt{g} \, \phi   \left( R  + \frac{2 \lambda }{L_2^2}\right),  \label{a5jt}
\end{align}
where,
\begin{equation}
\lambda = \frac{1}{r_h^{5/2} } \left(1 - \frac{2r_h^2}{L^2} \right)^2 \left(1 + \frac{r_h^2}{L^2} \right).
\end{equation}
Now, the constant Weyl rescaling,
\begin{equation}
\label{w5}
g_{\alpha \beta} \rightarrow \frac{1}{\lambda} g_{\alpha \beta},
\end{equation}
brings the action to the form (\ref{JTact}) with $\widetilde{G}$ given by (\ref{valtg5}). It is worth pointing out that since we are  expanding about the attractor eq.(\ref{kads52S0}), eq.(\ref{attracphi}),  there  is no term  in eq.(\ref{a5jt})  which is linear in $\sigma$.


\section{Rotating Dyonic Black Holes in Four Dimensions in Asymptotically Anti-de Sitter Spacetime}
\label{app-rot4d}

In this appendix, we elaborate on the four-dimensional set-up discussed in section \ref{sec-4drbf}. The action,  metric, gauge field and the various parameters are described in eqs.(\ref{dac4}-\ref{m4}).

The expressions for entropy and temperature are given respectively by,
\begin{align}
S &= \frac{\pi}{\Xi G_{4} } (r_+^2 + a^2 ), \label{kads-ent} \\
T &= \frac{(r_+^2 -a^2 - q_e^2 - q_m^2) +\frac{r_+^2}{L^2}( 3r_+^2 +a^2)}{4\pi r_+ (r_+^2 + a^2)}. \label{kads-temp}
\end{align}
Here $r_+$ is the location of the outer horizon, $\Delta_r (r_+)=0$. The angular velocity at the horizon is given by,
\begin{equation}
\label{omh4}
\Omega_H \equiv \omega(r_+) = \frac{a \Xi}{r_+^2 +a^2}.
\end{equation} 
As in the five-dimensional case, the angular velocity which is thermodynamically conjugate to the angular momentum \cite{Gibbons:2004ai} is given by,
\begin{equation}
\label{omhi4}
\Omega_H^\infty = a  \frac{\left( 1 + \frac{r_+^2}{L^2} \right)}{r_+^2 + a^2}.
\end{equation}

Let us now examine the black hole around extremality. Denoting all extremal quantities with a subscript $0$ and defining 
\begin{equation}
q_0^2 \equiv q_{e0}^2 + q_{m0}^2, \label{defq4}
\end{equation} 
the mass parameter $m_0$ and and the angular momentum parameter $a_0$ are parametrically related as,
\begin{align}
 m_{0} &= r_h \frac{ \left( 1 + \frac{2r_h^2 - q_0^2}{L^2} + \frac{r_h^4}{L^2 } \right) }{\left(1 -\frac{r_h^2}{L^2} \right)}, \label{m04} \\
a_{0} &= r_h \left( \frac{1+ \frac{3r_h^2}{L^2} - \frac{q_0^2}{r_h^2}}{1- \frac{r_h^2}{L^2}} \right)^{1/2}, \label{a04}
\end{align}
where $r_h$ is the radial location of the extremal horizon.

For a concise presentation of the later results, it is useful to express the mass and charge parameters as functions of the extremal radius and the angular momentum parameter. So we have,
\begin{align}
 m_{0} &= r_h \left(1 + \frac{a_0^2 + 2r_h^2}{L^2} \right), \label{m2} \\
q_{0} &= \sqrt{ r_h^2 \left(1+ \frac{3r_h^2}{L^2} \right) - a_0^2 \left(1- \frac{r_h^2}{L^2} \right)}. \label{q2}
\end{align}

We are interested in making the black hole slightly non-extremal by adding a small amount of mass, but keeping the angular momentum and the electric and magnetic charges unchanged. This necessitates that we take,
\begin{align}
m &= m_{0} + \delta m, \label{masd4} \\
a &= a_0\left( 1 - \frac{ \Xi_0}{ \left( 1 + \frac{3a_0^2}{L^2}  \right) } \frac{\delta m}{m_0} \right), \label{adef4} \\
q_e &= q_{e0} \left(1 + \frac{2 a_0^2}{L^2 + 3 a_0^2 } \frac{\delta m}{m_0} \right), \label{qed4} \\
q_m &=q_{m0} \left(1 + \frac{2 a_0^2}{L^2 + 3 a_0^2 } \frac{\delta m}{m_0} \right), \label{qmd4}
\end{align}
Then, the outer horizon is shifted to,
\begin{equation}
\label{kads-sh}
r_+ =  r_h + \delta_0 + \mathcal{O} \left( \delta_0^2 \right),
\end{equation}
where
\begin{equation}
\label{del04}
\delta_0^2 = \frac{2(r_h^2 + a_0^2) }{r_h \left(1 + \frac{3a_0^2}{L^2} \right)\left(1 + \frac{6r_h^2 + a_0^2}{L^2} \right) }  \delta m .
\end{equation}
The mass above extremality and the temperature are related as,
\begin{equation}
\label{delmT4}
\delta M =   \frac{ 2 \pi^2 r_h ( r_h^2 + a_0^2 ) }{G_{4} \Xi_0 \left( 1 + \frac{6r_h^2 + a_0^2}{L^2} \right) } T^2.
\end{equation}
The entropy above extremality then goes as,
\begin{equation}
\label{entT4}
\Delta S =  \frac{ 4 \pi^2 r_h ( r_h^2 + a_0^2 ) }{G_{4} \Xi_0 \left( 1 + \frac{6r_h^2 + a_0^2}{L^2} \right) } T. 
\end{equation}
Here,
\begin{equation}
\label{xi0}
\Xi_0 = 1 - \frac{a_0^2}{L^2}.
\end{equation}

We now look at the near-horizon limit of the metric. We make the  scalings as in \cite{Bardeen:1999px,Hartman:2008pb}, in a slightly more generalised form to accommodate deviations from extremality,
\begin{align}
t &\rightarrow {\lambda}^{-1} t, \nonumber \\
r - r_h &\rightarrow  \lambda (r-r_h), \nonumber \\
\varphi &\rightarrow \varphi - \mi   {\lambda}^{-1} \Omega_H t, \nonumber \\
m &=  m_{0} + \lambda^2 \delta m,  \label{scal}\\
a &= a_0\left( 1 - \lambda^2 \frac{ \Xi_0}{ \left( 1 + \frac{3a_0^2}{L^2}  \right) } \frac{\delta m}{m_0} \right), \nonumber\\
q_{(e,m)} &= q_{(e0, m0)} \left(1 + \lambda^2 \frac{2 a_0^2}{L^2 + 3 a_0^2 } \frac{\delta m}{m_0} \right) \nonumber.
\end{align}
Putting this into eq. (\ref{mt4}) and taking the limit $\lambda \to 0$ gives us the attractor solution,
\begin{align}
\md s^2 &= \frac{\rho_0^2 }{(r_h^2 +a_0^2) } \left(  \frac{(r-r_h)^2  - \delta_0^2}{L_2^2} \md t^2 + \frac{L_2^2}{(r-r_h)^2  - \delta_0^2} \md r^2 \right) + \frac{\rho_0^2 \Xi_0 }{(r_h^2 +a_0^2) \Delta_{\theta 0} } \Phi_0^2 \, \md \theta^2    \nonumber \\
&\quad + \frac{\Delta_{\theta 0} (r_h^2 + a_0^2) }{\Xi_0 \rho_0^2} \Phi_0^2 \sin^2 \theta  \left( \md \varphi -\mi \frac{2  r_h a_0 \Xi_0}{(r_h^2 + a_0^2)^2 }  (r-r_h) \, \md t \right)^2, \label{kam}
\end{align}
where,
\begin{align}
L_2^2 &= \frac{r_h^2 + a_0^2}{ \left( 1 + \frac{6r_h^2 + a_0^2}{L^2} \right) }, \label{defL24} \\
\rho_0^2 &= r_h^2 + a_0^2 \cos^2 \theta , \label{rat4} \\
\Phi_0^2 &=  \frac{r_h^2 + a_0^2}{\Xi_0}  , \label{phiat4} \\
\Delta_{\theta 0} &= 1 -  \frac{a_0^2}{L^2} \cos^2 \theta. \label{delat4}
\end{align}

The near-horizon gauge field takes the form, after removing a part going as $ k \, \md t$ by a gauge transformation,
\begin{align}
\hat{A}_\mu \, \md x^\mu &= \frac{q_{e0} }{\Xi_0 \rho_0^2} \left( -\mi \frac{ (r_h^2 - a_0^2 \cos^2 \theta) \Xi_0 }{r_h^2 + a_0^2} (r-r_h)\, \md t + a_0 r_h \sin^2 \theta \, \md \varphi \right) \nonumber  \\
&\quad + \frac{q_{m0} \cos \theta }{\Xi_0 \rho_0^2} \left( -\mi  \frac{2 a_0 r_h \Xi_0 }{r_h^2 + a_0^2} (r-r_h) \, \md t + (r_h^2 + a_0^2) \, \md {\varphi} \right). \label{kag}
\end{align}

Higher order corrections to the attractor metric and the gauge field can be obtained  by retaining terms of higher orders in $\lambda$. Even the first order corrections to the metric are rather complicated.

\subsection{Dimensional Reduction to Two Dimensions}

Next, we reduce the action (\ref{dac4}) with metric and gauge field ansatz to be of the form (\ref{metan4}) and (\ref{gfan4}), respectively. Note that we take 
$\Phi, A_\alpha, {\mathcal A}_\alpha$ to be only dependent on coordinates $x^\alpha$.

The resulting two-dimensional action then becomes, after including appropriate boundary terms for the canonical ensemble,
\begin{align}
I &= -\frac{1}{4 G_{4} } \int \md^2 x \, \sqrt{g } \left(R \Phi^2  + c_1\frac{\Phi_0}{\Phi} + c_2 \Phi_0 \Phi - c_3\frac{\Phi_0}{\Phi^3}   +  \frac{\Phi^3}{\Phi_0} ( c_4 \Phi^2 + c_5 ) F_{\alpha \beta} F^{\alpha \beta}     \right) \nonumber \\
&\quad  -\frac{1}{4 G_{4} } \int \md^2 x \, \sqrt{g }  \,  \frac{\Phi^3}{\Phi_0} \left(c_6 \mathcal{F}_{\alpha \beta} \mathcal{F}^{\alpha \beta} + c_7  \mathcal{F}_{\alpha \beta}  F^{\alpha \beta} \right)  - \frac{1}{2 G_{4} } \int \md x\, \sqrt{\gamma} \, \Phi^2 K   \label{euac2} \\
&\quad + \frac{1}{ G_{4} }  \int \md x\, \sqrt{\gamma} \,  \frac{\Phi^3}{\Phi_0} n_\alpha \left(  ( c_4 \Phi^2 + c_5 )  F^{\alpha \beta} A_\beta +  c_6   \mathcal{F}^{\alpha \beta} \mathcal{A}_{\beta} +    \frac{c_7 }{2} (F^{\alpha \beta} \mathcal{A}_{\beta} + \mathcal{F}^{\alpha \beta} A_\beta  )  \right) .\nonumber
\end{align}
where,
\begin{align}
c_1 &=  \frac{1}{\Xi_0}\left( 1 -\frac{r_h^2}{L^2} (3+ 2\chi^2) \right) + \frac{1}{\Xi_0 \chi} \left( (1-\chi^2) + \frac{r_h^2}{L^2} (3+\chi^2) \right) \tan^{-1} \chi,  \label{c1} \\
c_2 &= \frac{2}{L^2} \frac{3+\chi^2}{1+\chi^2}, \label{c2}\\
c_3 &= \frac{1}{\Xi_0^2 \chi (1+\chi^2)} \left( q_0^2 (1+\chi^2)^2 \tan^{-1} \chi - (q_{e0}^2 - q_{m0}^2 )\chi (1-\chi^2) \right), \label{c3} \\
c_4 &= \frac{(1+\chi^2)^2}{8\chi^3 \Xi_0} \left[ \left(1 +\frac{3r_h^2}{L^2} \right) \chi - \left( 1 - \chi^2 + \frac{r_h^2}{L^2} ( 3 + \chi^2)  \right) \tan^{-1} \chi \right], \label{c4} \\
c_5 &= \frac{(1+\chi^2)}{8 \chi^3 \Xi_0^2} \left[ \Big(q_{m0}^2 (1+\chi^2)^2 + q_{e0}^2 (3-2\chi^2 + 3\chi^4) \Big) \tan^{-1} \chi - (3 q_{e0}^2 + q_{m0}^2) \chi (1-\chi^2) \right], \label{c5} \\
c_6 &= \frac{(1+\chi^2)}{\chi} \tan^{-1} \chi, \label{c6} \\
c_7 &= \frac{q_{e0} (1+\chi^2) [ \chi - (1-\chi^2) \tan^{-1} \chi] }{\chi^2 \Xi_0}. \label{c7}
\end{align}

Here, we have defined,
\begin{equation}
\label{defchi}
\chi = \frac{a_0}{r_h},
\end{equation}
and have used the standard notation for field strength,
\begin{equation}
\label{defF4}
F_{\alpha \beta} = \partial_\alpha A_\beta - \partial_\beta A_\alpha, \quad \mathcal{F}_{\alpha \beta} = \partial_\alpha \mathcal{A}_\beta - \partial_\beta \mathcal{A}_\alpha.
\end{equation}
In the two-dimensional theory, we have two $\mathrm{U}(1)$ gauge fields. The equations of motion for $A_\alpha$ and $\mathcal{A}_\alpha$, give respectively,
\begin{align}
\frac{\Phi^3}{\Phi_0} ( c_4 \Phi^2 + c_5 ) F_{\alpha \beta} + c_7 \frac{\Phi^3}{2\Phi_0} \mathcal{F}_{\alpha \beta} &= Q \sqrt{g} \, \epsilon_{\alpha \beta}, \label{eom-A} \\
\frac{ c_6 \Phi^3}{\Phi_0} \mathcal{F}_{\alpha \beta} + c_7 \frac{\Phi^3}{2\Phi_0} F_{\alpha \beta} &= \mathcal{Q} \sqrt{g} \, \epsilon_{\alpha \beta}. \label{eom-A2}
\end{align}
It is easy to see from the attractor values (\ref{gf4r}, \ref{gf4r2}) that these charges are nothing but the extremal values of the angular momentum and the electric charge:
\begin{equation}
\label{qq}
Q = G_{4} J_0, \quad \mathcal{Q} = Q_{e0},
\end{equation}
as one would expect. As suggested by the notation, the attractor value of the field $\Phi$ is given by $\Phi_0$ and the attractor geometry is an $\mathrm{AdS}_2$ with,
\begin{equation}
\label{kar2}
R = - \frac{2}{L_2^2}.
\end{equation}

Following the general procedure discussed in \S\ref{ss-jtrot}, and expanding $\Phi$ about its attractor value $\Phi_0$, eq.(\ref{defphi}), eq.(\ref{phiat4}), we obtain the bulk action of the JT model,

\begin{align}
I =  -\frac{(r_h^2 + a_0^2)}{2 G_{4} \left( 1 - \frac{a_0^2}{L^2} \right)}  \int \md^2 x \, \sqrt{g} \, \phi \left(R+ \frac{2}{L_2^2} \right)  
\label{JT4}
\end{align}
From the above, we identify the two-dimensional Newton's constant in the JT model to be given by eq.(\ref{knag}).

From the full metric (\ref{mt4}), we get the field $\Phi$ to be,
\begin{equation}
\label{nphi}
\Phi^2 =  \frac{1}{\Xi} \sqrt{\frac{\Sigma}{\Delta_\theta}}.
\end{equation}
At the horizon in the extremal case, $\Delta_r$ which appears in $\Sigma$ has a second order zero. As a result we can set $\Delta_r$ to vanish in obtaining the 
linear variation of $\Phi$ away from the horizon. 
This gives, 
\begin{equation}
\label{phie}
\Phi = \Phi_0 \left( 1 + \frac{r - r_h}{L_2^2 \mathcal{J}} \right) = \Phi_0 \left( 1 + \frac{1}{ \mathcal{J} z} \right),
\end{equation}
where $\mathcal{J}$ is as defined in eq.(\ref{knaj}).

By taking certain limits of the parameters, one can easily obtain different special cases:
\begin{itemize}
\item The asymptotically flat spacetime limit is obtained by taking $L \to \infty$
\item The purely rotating limit is obtained by taking $q_{e0}, q_{m0} \to 0$ and $\mathcal{A}_\alpha \to 0$.
\item The purely magnetic limit (with rotation) is obtained by taking $q_{e0} \to 0$ and $\mathcal{A}_\alpha \to 0$.
\item The purely electric limit (with rotation) is obtained by taking $q_{m0} \to 0$.
\item The non-rotating limit is obtained by taking $a_0 \to 0$ and $A_\alpha \to 0$.
\end{itemize}


\bibliographystyle{JHEP}
\bibliography{ref}
\end{document}